\documentclass[12pt,preprint]{aastex}

\newcommand{\mathsym}[1]{{}}
\newcommand{\unicode}[1]{{}}

\newcommand{\beq}[1]{\begin{equation}\label{#1}}
\newcommand{\eeq}{\end{equation}}
\newcommand{\bnq}{\[}
\newcommand{\enq}{\]}

\newcommand{\bea}[1]{\begingroup\setlength\arraycolsep{1.4pt}\begin{eqnarray}\label{#1}}
\newcommand{\eea}{\end{eqnarray}\endgroup}
\newcommand{\bna}{\begingroup\setlength\arraycolsep{1.4pt}\begin{eqnarray*}}
\newcommand{\ena}{\end{eqnarray*}\endgroup}

\newcommand{\equ}[1]{{\rm Eq.(\ref{#1})}}

\newcommand{\np}{\noindent}

\shorttitle{The obliquity of Mercury}
\shortauthors{Noyelles \& Lhotka}

\usepackage{wasysym}

\begin{document}


\title{The influence of orbital dynamics, shape and tides on the obliquity of Mercury}


\author{Beno\^it Noyelles \& Christoph Lhotka\footnote{Now at Dipartimento di Matematica -- Universit\`a degli Studi di Roma Tor Vergata -- Via della Ricerca Scientifica, 1, 00133 Roma -- Italy}} \affil{University
of Namur, Dpt of Mathematics and Namur Center for Complex Systems (NAXYS), 8
Rempart de la Vierge, B-5000 Namur, Belgium, tel: +3281724940 fax: +3281724914}\email{benoit.noyelles@unamur.be,lhotka@mat.uniroma2.it}



\begin{abstract}

	\par Earth-based radar observations of the rotational dynamics of Mercury \citep{mpshgjygpc2012}
	combined with the determination of its gravity field by MESSENGER \citep{szpshlmnpmjtprght2012} give clues
	on the internal structure of Mercury, in particular its polar moment of inertia $C$, deduced from the obliquity $(2.04\pm0.08)$ arcmin.
	
	\par The dynamics of the obliquity of Mercury is a very-long term motion (a few hundreds of kyrs), based on the regressional 
	motion of Mercury's orbital ascending node. This paper, following the study of \citet{nd2012}, aims at first giving
	initial conditions at any time and for any values of the internal structure parameters for numerical simulations, 
	and at using them to estimate the influence of usually neglected parameters on the obliquity, like $J_3$, the Love
	number $k_2$ and the secular variations of the orbital elements. We use, for that, averaged representations of the 
	orbital and rotational motions of Mercury, suitable for long-term studies.
	
	\par We find that $J_3$ should alter the obliquity by 250 milli-arcsec, the tides by 30 milli-arcsec, and the secular 
	variations of the orbital elements by 10 milli-arcsec over 20 years. The resulting value of $C$ could be at the most 
	changed from $0.346mR^2$ to $0.345mR^2$.

\end{abstract}





\section{Introduction}

\par The space missions MESSENGER (NASA) and BepiColombo (ESA / JAXA)
\citep{bgsslkf2007} are opportunities to have a better knowledge of Mercury, in
particular its rotational dynamics and its internal structure. The size of an
outer molten core can be inverted from the well-known Peale experiment
\citep{p1976,ppssz2002} in measuring separately the obliquity of Mercury,
yielding the polar momentum of inertia $C$, and the longitudinal librations,
yielding the polar momentum of the mantle $C_m$. This last assertion is based
on the assumption that the longitudinal librations of the mantle are decoupled
from the other layers, what can be doubtful if Mercury has a significant inner
core \citep{vd2011,vrby2012}. But this does not affect the determination of
$C$.

\par The determination of $C$ from the obliquity is based on the assumptions
that Mercury is at the Cassini State 1, and that it behaves as a rigid body
over the timescales relevant for the variations of the obliquity, i.e. a few
hundreds of years, due to the regressional motion of Mercury's orbital nodes.
The Cassini State 1 is a dynamical equilibrium in which Mercury is in a 3:2
spin-orbit resonance \citep{pd1965,c1965}, and the small free librations around
the equilibrium have been damped with enough efficiency so that the obliquity
of Mercury can be considered as an equilibrium obliquity. In the following, this obliquity will be denoted as $\epsilon$ and is the angle between the normal to the orbit and the angular momentum of Mercury. We will also consider the \emph{inertial obliquity} K, defined as the angle between the normal to an inertial reference plane and the angular momentum. \citet{p2005} estimated the damping timescale to be of the order of $10^5$ years for the free
librations and the free precession of the spin, in considering the tidal
dissipation and the core-mantle friction. The presence of Mercury near
the Cassini State 1 has been confirmed by Earth-based radar observations
\citep{mpjsh2007,mpshgjygpc2012}, giving an obliquity of $2.04 \pm 0.08$
arcminutes. The variations of the obliquity can be considered as small
and adiabatic \citep{bc2005,p2006,br2007,dl2008}.

\par It is commonly accepted that the obliquity $\epsilon$ should be inverted using Peale's formula \citep{p1969}, that reads as \citep{ym2006}:

\begin{equation}
	\label{eq:peale}
	\epsilon=-\frac{C\dot{\ascnode}\sin i}{C\dot{\ascnode}\cos i+2nmR^2\left(\frac{7}{2}e-\frac{123}{16}e^3\right)C_{22}-nmR^2\left(1-e^2\right)^{-3/2}C_{20}},
\end{equation}
where $n$ is the orbital mean motion of Mercury, $m$ its mass, $R$ its mean radius, $e$ its orbital eccentricity, $C_{20}=-J_2$ and $C_{22}$ 
are the most 2 relevant coefficients of the gravity field of Mercury, and $i$ and $\dot{\ascnode}$ are the inclination and nodal precession rate 
of the orbit of Mercury with respect to a Laplace Plane. All these quantities are assumed to be constant. The Laplace Plane is a reference plane, determined so as to minimize the variations of the inclination $i$. There are several ways to define the Laplace Plane \citep{ym2006,br2007,dndl2009}, and so $i$ and $\dot{\ascnode}$ depend on the chosen definition.

\par In order to bypass the uncertainty on the Laplace Plane, \citet{nd2012} proposed a Laplace Plane-free study of the obliquity, using averaged 
equations of the rotational motion, averaged variations of the eccentricity, inclination and the associated angles, and a quasi-periodic representation of 
these quantities. This yields a quasi-periodic representation of the equilibrium obliquity that is very close to the Cassini State over the domain of
validity of the JPL DE 406 ephemerides \citep{s1998}, i.e. 6,000 years between JED 0625360.50 (-3000 February 23) and 2816912.50 (+3000 May 06). The inertial reference frame is the ecliptic at J2000. This approach allows to consider the small variations of the orbital elements.

\par The spacecraft MESSENGER currently orbiting around Mercury has recently given us accurate gravity coefficients (Tab.\ref{tab:smith}) \citep{szpshlmnpmjtprght2012}. This is the opportunity to refine our model of the obliquity and to consider usually neglected effects like the tides or higher order gravity coefficients. \\

\begin{table}[ht]
	\centering
	\caption{The gravity field coefficients of Mercury derived from MESSENGER data \citep{szpshlmnpmjtprght2012}. These are unnormalized coefficients 
	while Smith et al. give them normalized.\label{tab:smith}}
	\begin{tabular}{l|r}
		\hline
		$C_{20}=-J_2$ & $(-5.031\pm 0.02)\times 10^{-5}$ \\
		$C_{21}$ & $(-5.99\pm 6.5)\times 10^{-8}$ \\
		$S_{21}$ & $(1.74\pm 6.5)\times 10^{-8}$ \\
		$C_{22}$ & $(8.088\pm 0.065)\times 10^{-6}$ \\
		$S_{22}$ & $(3.22\pm 6.5)\times 10^{-8}$ \\
		$C_{30}=-J_3$ & $(-1.188\pm 0.08)\times 10^{-5}$ \\
		$C_{40}=-J_4$ & $(-1.95\pm 0.24)\times 10^{-5}$ \\
		\hline
	\end{tabular}
\end{table}	

\par The dynamics of the obliquity is a long-term dynamics, since it is ruled by the precessional motion of Mercury's orbital node, its period being of the order of 300 kyr. That is the reason why we should average the equations over the short-period perturbations, the period associated being of the order of the year. In previous studies on the subject, only the main perturbative effects were taken into account. In the present work we therefore aim to investigate also the influence of these additional perturbative terms on the obliquity of Mercury: the influence of higher order terms in eccentricity, higher order gravity harmonics, the influence of tides, in both averaged and unaveraged models of the spin-orbit interactions of Mercury.

\par We first present how we average the equations ruling the rotational dynamics of Mercury (Sect.\ref{sec:dynmod}). This averaging is more accurate than the one given in \citep{nd2012}, i.e. it is done in higher order in eccentricity. Then we explain how we derive 2 new formulae for the obliquity, one analytically (Sect.\ref{sec:analytical}), with an approach slightly different than Peale's, and one numerically (Sect.\ref{NUMS}). Then the reliability of these formulae are tested and additional effects are discussed (Sect.\ref{sec:influence}). The influence on the interpretation of the internal structure of Mercury is finally addressed (Sect.\ref{sec:invert}).

\section{The dynamical model\label{sec:dynmod}}

\np This section aims at deriving the equations of the long-term rotation of Mercury. For that we start from the exact equations of the problem, and we expand the potential with respect to the orbital parameters (eccentricity and inclination) and to the spherical harmonics of the gravity field of Mercury (Tab.\ref{tab:smith}). This expansion induces the apparition of fast sinusoidal perturbations, that disappear after averaging. Such a calculation is already present in \citep{nd2012}, but the gravity field of Mercury is there considered only up to the second order, and the expansion in eccentricity / inclination limited to the degree 3.

\np The basic model behind the spin-orbit dynamics of the Sun-Mercury system is
given in terms of the Hamiltonian  (\cite{dl2008}):

\beq{Eq0}
\mathcal{H}=\mathcal{H}_K+\mathcal{H}_R-(G_c M m)V_G,
\eeq

\np where $G_c$ is the gravitational constant, $M$ is the mass of the Sun and
\(m\simeq 1.6\cdot 10^{-7} M\) is the mass of Mercury. The symbol \(\mathcal{H}_K\)
labels the unperturbed Kepler problem:

\beq{Eq0a}
\mathcal{H}_K=-\frac{m^3\mu ^2}{2L_o^2} \ ,
\eeq

\np with the constant $\mu$ given by \(\mu =G_c(M+m)\), and \(L_o=m\sqrt{\mu a}\)
is the orbital angular momentum, which depends on the semi-major axis of Mercury
\(a\simeq 0.387 AU\). \(\mathcal{H}_R\) defines the free rotational motion of Mercury:

\beq{Eq0b}
\mathcal{H}_R=\frac{1}{2}\left(G^2-L^2\right)\left(\frac{\sin ^2(l)}{A}+
\frac{\cos ^2(l)}{B}\right)+\frac{L^2}{2C} \ ,
\eeq

\np where \(A\leq B<C\) are the principal moments of inertia, \(G\) is the norm of
the angular momentum $\vec{G}$, \(L=G \cos (J)\) is the projection of \(\vec{G}\) onto the
polar figure axis of Mercury, with the angle \(J\) usually called the wobble,
and \(l\) is the conjugated angle to the action \(L\). Physically, $l$ represents the precession of the geometrical polar axis (or figure axis) of 
Mercury about the angular momentum, and $J$ is the amplitude of this motion.

\subsection{General form of the potential}

\np The gravitational interaction of the orbital and rotational dynamics is given by
the potential \(V_G\). It can be expanded into spherical harmonics, and takes
the form, after \citep{Cun1970,bf1990}:

\beq{Eq1}
V_G=\sum _{n=0}^{\infty } \sum _{m=0}^n \frac{R^n}{r^{n+1}}
P_{nm}(\sin  \varphi )\left(C_{nm}\cos  (m \lambda )+S_{nm}
\sin (m \lambda)\right) \ ,
\eeq

\np where \(R\simeq 2439.7 km\) \citep{aabcccfhhknossttw2011} is the radius of Mercury, \(r\) is the
distance of the center of mass of Mercury from the center of mass of the Sun,
\(P_{nm}=P_{nm}(u)\) are the associated Legendre polynomials, which are defined
in terms of the standard Legendre polynomials \(P_n=P_n(u)\) by:

\bnq
P_{nm}(u)=\left(1-u^2\right)^{m/2}\frac{d\, P_n(u)}{d\, u^m}.
\enq

\np The angles \((\varphi ,\lambda )\) are latitude and longitude with \(\varphi
\in (-90^{\circ},90^{\circ})\), \(\lambda \in (0,360^{\circ})\), and
the \(C_{nm}\), \(S_{nm}\) are the Stokes coefficients, with
\(m\leq n\) and \(n,m\in \mathbb{N}\). The standard convention is to define
\(C_{00}=1\) and \(S_{n0}=0\). The notation \(J_n=-C_{n0}\) is also used for
the remaining zonal terms, with \(m=0\). In addition, by the proper choice of
the coordinate system through the center of mass of Mercury, the first order
coefficients \(C_{10}\), \(C_{11}\) and \(S_{11}\) vanish, the same is true for
\(C_{21}=S_{21}=0\) if the axes of figure are aligned with the main axes of
inertia. The effect of the perturbation on the rotation is therefore
proportional to the size of the remaining zonal \((m=0)\), tesseral \((m<n)\)
and sectorial \((m=n)\) terms. The remaining second order
coefficients, \(C_{20}=-J_2\) and \(C_{22}\), can be related to the principal
moments of inertia by:

\bnq
J_2m R^2=C-\frac{(A+B)}{2}, C_{22}m R^2=\frac{B-A}{4} \ .
\enq

\np It is a common practize to introduce the normalized polar moment of inertia
\(c\) through the additional equation \(C=c m R^2\).

\np The contributions \(\mathcal{H}_K,\mathcal{H}_R\) and \(V_G\) are given in different reference
frames. Let us denote by \(e_0\)
the inertial, by \(e_1\) the orbital, by \(e_2\) the spin and by \(e_3\) the
figure frame of references, respectively. In the following, the inertial frame will be either a Laplace frame, minimizing the variations of 
the orbital inclination, or the ecliptic at J2000.0. These choices of course affect the definitions of the inclination $i$, of the 
ascending node $\ascnode$, and of variables of rotation $g$ and $h$, defined later.
To match \(\mathcal{H}_K,\mathcal{H}_R\) and \(V_G\) we aim to express them in the inertial frame
\(e_0\). For this reason we introduce the unit vector
\(\left(\hat{x},\hat{y},\hat{z}\right)\), pointing from the center of mass of
Mercury to the one of the Sun, which is defined in the body frame \(e_3\),
in terms of the longitude $\lambda$ and the latitude $\varphi$ by the relations:

\beq{Eq2}
\hat{x}=\cos  \varphi  \cos  \lambda , \ \
\hat{y}=\cos  \varphi  \sin  \lambda , \ \ 
\hat{z}=\sin  \varphi \ .
\eeq

\np Together with the definition of \(P_n\) in terms of the sum

\bnq
P_n(u)=\frac{1}{2^n}\sum _{k=0}^{[n/2]}
\frac{(-1)^k(2n-2k)!}{k!(n-k)!(n-2k)!}u^{n-2k}
\enq

\np we find for \(u=\sin  \varphi\) the explicit form for \(P_{nm}\), which are given by:

\beq{Eq3}
P_{nm}(\sin  \varphi )=\frac{1}{2^n}\cos ^m(\varphi )
\sum _{k=0}^{[(n-m)/2]} \frac{(-1)^k(2n-2k)!}
{k!(n-k)!(n-2k-m)!}\sin ^{n-2k-m}(\varphi) \ .
\eeq

\np Together with the formulae by Vieta (see e.g. \citet{h2001})

\bnq
\begin{array}{c}
 \sin  \\
 \cos  \\
\end{array}
(m \lambda )=\sum _{k=0}^m \left(
\begin{array}{c}
 m \\
 k \\
\end{array}
\right)\cos ^k\lambda  \sin ^{m-k}\lambda  
\begin{array}{c}
 \sin  \\
 \cos  \\
\end{array}
\left(\frac{1}{2}(m-k)\pi \right)
\enq

\np we find from \equ{Eq1}, \equ{Eq2} and \equ{Eq3}:

\beq{Eq4}
V_G=\frac{1}{r}\sum _{n=0}^{\infty } \sum _{m=0}^n 
\left(\frac{R}{r}\right)^n\left(C_{nm}
\mathfrak{C}_{nm}+S_{nm}\mathfrak{S}_{nm}\right) \ ,
\eeq

\np where the $\mathfrak{C}_{nm}$, $\mathfrak{S}_{nm}$ are now
functions of $(\hat x, \hat y, \hat z)$ only. Note, that using
the property $\hat x^2+\hat y^2+\hat z^2=1$ they can be
expressed in different forms. We provide the main terms that
we are going to use in the present study in Sect.~\ref{SPOT}.

\subsection{Matching the reference frames}

\np To express the unit vector \(\left(\hat{x},\hat{y},\hat{z}\right)\) in \(e_0\)
we make use of the usual Andoyer angles \((l,g,h)\) together with the
angles \((J,K)\); the angle \(J\) was already defined above, and the angle \(K\)
enters the projection of \(\vec{G}\) on the inertial \(z\)-axis by \(H=G \cos (K)\).
As already stated, we call $K$ the \emph{inertial obliquity}. The angle $h$ is
a node representing the precession of the angular momentum with respect to the orbital plane,
and $g$ can be seen as the spin angle.

\np Let us denote by $R_1,R_2,R_3$ the rotation matrices around the
$x,y,z$ - axes, respectively. The unit vector \(\left(\hat{x},
\hat{y},\hat{z}\right)\), given in \(e_3\), can be expressed in the inertial
frame \(e_0\) by

\bnq
\left(\hat{x},\hat{y},\hat{z}\right)_{e_3}=R_M\cdot (\cos (f),\sin (f),0) \ ,
\enq

\np where $f$ is the true anomaly, and we introduced the rotation matrix $R_M$ 
being of the form:

\bnq
R_M=R_3(-l)R_1(-J)R_3(-g)R_1(-K)R_3(-h)R_3(\ascnode )R_1(i)R_3(\omega ) \ .
\enq

\np Together with the relations

\bnq
\frac{a}{r}=1+2\sum _{\nu =1}^{\infty } J_{\nu }(\nu  e) \cos (\nu  \mathcal{M}) \ ,
\enq

\bnq
\cos (f)=2\frac{1-e^2}{e}\sum _{\nu =1}^{\infty } J_{\nu }(\nu  e) \cos (\nu  \mathcal{M})-e \ ,
\enq

\bnq
\sin (f)=2\sqrt{1-e^2}\sum _{\nu =1}^{\infty } \frac{d J_{\nu }(\nu  e)}{d e}
\frac{\sin (\nu  \mathcal{M})}{\nu } \ ,
\enq

\np where \(J_{\nu }\) are the Bessel functions of the first kind and $\mathcal{M}$ is the mean anomaly, we are able to
express the potential \equ{Eq4} in terms of the rotational and orbital elements
only:

\beq{POT}
V_G=V_G(l,g,h,J,K,a,e,i,\omega ,\mathcal{M}).
\eeq

\np It can also be expressed in terms of suitable action angle variables
\(\left(l_i,L_i\right)\), with \(i=1,\ldots ,6\), using the set of modified
Andoyer variables

\bea{AND}
\left(l_1,\ l_2,\ l_3\right)&=&(l+g+h,\ -l,\ -h) \ , \nonumber \\
\left(L_1,\ L_2,\ L_3\right)&=&(G,\ G-L,\ G-H)=(G,G(1-\cos(J)),\  G(1-\cos(K)))
\eea

\np for the rotational motion, and by making use of the classical Delaunay
variables

\bea{DEL}
\left(l_4,\ l_5, \ l_6\right)&=&(\mathcal{M}, \ \omega, \ \ascnode ) \ , \nonumber\\
\left(L_4, \ L_5, \ L_6\right)&=&\left(L_o, \ L_4\sqrt{1-e^2}, \ L_5 \cos (i)\right)
\eea

\np for the orbital dynamics. In this setting the potential can be written in the
form \(V=V(l,L)\) with \(l=\left(l_1,\ldots ,l_6\right)\) and
\(L=\left(L_1,\ldots ,L_6\right)\).

\subsection{Simplifications and assumptions}

\np The rotation period of Mercury, about
\(T_r=58.6 d\), and the orbital period around the Sun, \(T_o\simeq 87.9 d\)
lie close to the \(3:2\) resonance \(\left(2T_o\simeq 3T_r\right)\). It is thus
desirable to find a much simpler dynamical model, which reproduces the qualitative dynamics
close to the resonance. We first introduce the change of coordinates

\bnq
S_{3:2}:(l,L){\mapsto}(\sigma ,\Sigma )
\enq

\np with \(\sigma =\left(\sigma _1,\ldots ,\sigma _6\right)\), \(\Sigma
=\left(\Sigma _1,\ldots ,\Sigma _6\right)\) defined by the generating function
\(S_{3:2}\) of the second kind

\bnq
S_{3:2}=\Sigma _1\left(l_1-\frac{3}{2}l_4-l_5-l_6\right)+
\Sigma _2l_2+\Sigma _3\left(l_3+l_6\right)+
\Sigma _4l_4 +\Sigma _5l_5+\Sigma _6l_6  \ .
\enq

\np In this setting the relevant resonant dynamics can be easily described in terms of
the variables

\bea{RES}
\sigma _1&=&l_1-\frac{3}{2}l_4-l_5-l_6 \ , \ 
\sigma _2=l_2 \ , \ 
\sigma _3=l_3+l_6 \ , \nonumber \\
\Sigma _1&=&L_1 \ , \ 
\Sigma _2=L_2 \ , \ 
\Sigma _3=L_3 \ ,
\eea

\np while the remaining variables become:

\bna
2\Sigma _4=2L_4+3\Sigma _1 \ , \
\Sigma _5=L_5+\Sigma _1 \ , \ 
\Sigma _6=L_6+\Sigma _1-\Sigma _3 \ ,
\ena

\np and $\sigma _i=l_i$ with  $i=4,5,6$.  In our first approach we aim to construct
a simple resonant model, valid only close to exact resonance, by making use of
the following assumptions, that we also justify briefly:

\begin{itemize}

\item[i)] we neglect the wobble motion of Mercury, i.e. we assume that the 
figure polar axis is the rotation axis; we therefore set \(J=0\),
which implies \(L_2=0\) and reduces \equ{Eq0b} to:

\beq{Eq0c}
\mathcal{H}_R=\frac{\Sigma _1{}^2}{2C}
\eeq

This motion should in fact induce a deviation of about only 1 meter of the spin pole from the geometrical North pole \citep{ndl2010}.

\item[ii)] we neglect the effect of the rotation on the orbital dynamics, i.e.
we investigate the dynamics on the reduced phase space

\bnq
\frac{d \sigma _i}{d t}=\frac{\partial \mathcal{H}}{\partial \Sigma _i} \ , \ 
\frac{d \Sigma _i}{d t}=-\frac{\partial \mathcal{H}}{\partial \sigma _i} \ , 
\enq

with $i=1,3$, therefore assume that the orbital parameters \(a,e,i,\omega
,\ascnode ,\mathcal{M}\) are known quantities, and $\Sigma_i=\Sigma_i(t),
\sigma_i=\sigma_i(t)$ with $i=4,5,6$ act as external time-dependent parameters
on the dynamics of the reduced phase space.

The assumption is valid since the effect of the rotation of Mercury on its 
orbit is much smaller compared to the perturbations due to the other planets.
The reason is that the energy associated with the rotational dynamics is negligible with respect to the orbital energy.

\item[iii)] we neglect short periodic effects\footnote{ The effects within time
scales, which are smaller than the revolution period of Mercury around the
Sun.} and replace the potential \(V_G\) by its average over the mean anomaly of
Mercury \(\mathcal{M}=l_4=\sigma _4\), which we denote by $\langle V\rangle$,
in short:

\bnq
\langle V\rangle \equiv \left\langle V_G\right\rangle _{\sigma _4}=
\frac{1}{2\pi }\int _0^{2\pi }V_G(\sigma ,\Sigma )d\sigma _4 \ .
\enq

As a result the averaged potential $\langle V\rangle$ becomes independent
of $\sigma_4$, or the mean anomaly $\mathcal{M}$, and thus the conjugated action
$\Sigma_4$ becomes a constant of motion, and as a consequence the semi-major
axis $a$ is assumed to be constant too.

The assumption preserves the qualitative aspects of the dynamics since the 
orbital and rotational periods of Mercury are short compared to the 
periods of revolution of the remaining nodes. By averaging theory we maintain 
the qualitative aspects of the dynamics also in the averaged model. \citet{dnrl2009} have estimated the influence of the short-period oscillations to be smaller than 20 milli-arcsecond on the obliquity.

\item[iv)] we assume the presence of a perturbation leading to an additional
precession of the nodes with constant precession rates, say \(\dot{\omega },
\dot{\ascnode }\neq 0\). In this setting we find for the remaining phase state
variables, connected to the orbital motion, $\sigma_5$ and $\sigma_6$:

\bnq
\frac{d \sigma _5}{d t}=\dot{\omega } \ , \ \frac{d \sigma _6}{d t}=\dot{\ascnode } \ .
\enq
These are the only remaining time dependent variables, their 
frequencies being constant.

This assumption has to be discussed: the main influence of the 
perturbations of the other planets that are important for the long-term 
dynamics of the rotational motion of Mercury are the secular perturbations 
of the nodes. In a first order approximation we therefore implement this 
cumulative effect by a linear precession of the nodes. Our first assumption
$\dot{\ascnode}=const$ is also in agreement with one of the definitions of 
the Laplace plane. If we therefore define $\dot{\ascnode}$ with respect to the
Laplace plane we optimize our results.
We also consider a constant precession rate $\dot{\omega}$.
This is a pretty good approximation if the reference plane is the ecliptic at
J2000, as given by the JPL HORIZONS website over 6,000 years. If we use
another reference plane like the Laplace plane, then the argument of the
pericenter $\omega_l$ is defined from a different origin. The two definitions of 
the pericentre result in very similar precession rates, the difference being due 
to the differential precession of the orbital plane with respect to these 
2 references planes, this is a small effect that we can safely neglect.

\end{itemize}

\np Note that since $\sigma _5=\sigma _5(t)=\dot{\omega} t+\omega_0$ and
	$\sigma _6=\sigma _6(t)=\dot{\ascnode}t+\ascnode_0$ the
time dependent generating function \(S_{3:2}\) leads to

\bnq
\frac{\partial S_{3:2}}{\partial t}=-\Sigma _1\dot{\omega }+
\left(\Sigma _3-\Sigma _1\right)\dot{\ascnode }
\enq

\np which we have to add to our new Hamiltonian. The final resonant model takes
the form

\beq{EQU}
\mathcal{H}_I=h_0-G_c M m \ \langle V\rangle  \ ,
\eeq

\np with the new fundamental part:

\beq{FUN}
h_0=\mathcal{H}_K+\mathcal{H}_R-\Sigma _1\dot{\omega }+\left(\Sigma _3-\Sigma _1\right)\dot{\ascnode }
\eeq

\np and where $\mathcal{H}_K$ transforms into the new expression:

\bnq
\mathcal{H}_K=-\frac{2m^3 \mu ^2}
{\left(2\Sigma _4-3\Sigma _1\right)^2} \ .
\enq

\np We derive from \equ{FUN} the unperturbed angular frequencies:

\bna
\dot{\sigma }_1&=&\frac{\partial h_0}{\partial \Sigma _1}=
\frac{\Sigma _1}{C}-\frac{3}{2}\frac{ m^3 \mu ^2}
{ \left(\Sigma _4-\frac{3}{2}\Sigma _1\right){}^3}-
\dot{\omega}-\dot{\ascnode }=\dot{l}_1-\frac{3}{2}n-\dot{\omega }-
\dot{\ascnode } \ , \nonumber \\
\dot{\sigma }_3&=&\frac{\partial h_0}{\partial \Sigma _3}=\dot{\ascnode }
\ena

\np since

\bnq
\dot{\sigma }_4=\frac{\partial h_0}{\partial \Sigma _4}=
\frac{m^3 \mu ^2}{\left(\Sigma _4-\frac{3}{2}\Sigma _1\right){}^3}\equiv n \ .
\enq

\np Here \(n\) is the mean motion of Mercury. For \(\dot{\omega }=\dot{\ascnode }=0\)
the spin-orbit resonance of Mercury translates into the commensurability:

\bnq
2\dot{\sigma}_1=2\dot{l}_1-3\dot{\sigma }_4 = 0 \ ,
\dot{\sigma }_3-\dot{\sigma _6} = 0 \ ,
\enq

\np while for \(\dot{\omega },\dot{\ascnode }\neq 0\) small frequency corrections due
to the potential \(\langle V\rangle\) have to be taken into account.

\subsection{Workout of the time dependent resonant model}\label{SPOT}

\np In the following discussion we limit our investigation to the contributions of
the averaged potential, in which the Stokes coefficients in Table~\ref{tab:smith}
are bigger than the threshold \(10^{-7}\), which turn out to be
 \(C_{20},C_{22},C_{30},C_{40}\). According to this simplification the averaged
potential \(\langle V\rangle\) can be split into the form:

\bnq
\langle V\rangle =C_{20}\left\langle V_{20}\right\rangle +
C_{22}\left\langle V_{22}\right\rangle +
C_{30}\left\langle V_{30}\right\rangle +
C_{40}\left\langle
V_{40}\right\rangle \ ,
\enq

\np where we used the notation \(\left\langle V_{nm}\right\rangle\) to indicate
the terms proportional to the coefficient \(C_{nm}\). For the ongoing
investigation we also need to separate the individual terms into purely
resonant terms, just depending on \(\sigma _1,\sigma _3\), and time dependent
resonant terms through the presence of the additional angles \(\sigma
_5=l_5=\omega ,\sigma _6=l_6=\ascnode\), with \(\omega =\omega (t), \ascnode
=\ascnode (t)\). These terms allow to see the variations of the obliquity
because of the variations of the precessional motion, while the time
independent terms contain only the mean precession rate of the node $\ascnode$.
We use the short-hand notation:

\beq{EqVt}
\left\langle V_{nm}\right\rangle=\left\langle
V_{nm}\right\rangle \left(\sigma _1,\sigma _3,l_5,l_6\right)=
f_{nm}\bigg(
\left\langle v_{nm}\right\rangle \left(\sigma_1,\sigma _3\right)+
\left\langle u_{nm}\right\rangle
\left(\sigma _1,\sigma _3,l_5,l_6\right)\bigg) \ ,
\eeq

\np where $f_{nm}$ denotes a common factor, $v_{nm}$ labels the time-independent 
resonant and constant terms, while $u_{nm}$ labels the time-dependent 
resonant terms, only.

\subsection*{The term \(\left\langle V_{20}\right\rangle\):}

\np The expression proportional to \(C_{20}\) in the unaveraged 
potential \(V_G \) takes the form:

\bnq
\frac{R^2}{r^3}\mathfrak{C}_{20}=\frac{1}{2}\frac{R^2}{r^3}
\left(3\hat{z}^2-1\right) \ ,
\enq

\np which becomes after the average over the mean anomaly, and expanded
up to $4$th order in the orbital eccentricity $e$:

\bnq
\left\langle V_{20}\right\rangle =-\frac{R^2\left(8+12e^2+15e^4\right)}
{256a^3}\left([1]_{20}+[2]_{20}\cos \left(\sigma _3\right)+
[3]_{20}\cos \left(2\sigma _3\right)\right)+O(e^5) \ .
\enq

\np The coefficients $[j]_{20}$, $j=1,2,3$ depend on $\Sigma_1$,
$\Sigma_3$ through the norm of the angular momentum $G$ and the inertial
obliquity $K$ (see \equ{AND}, \equ{RES}), as well as on the orbital
inclination $i$, and can also be found in the Appendix. Note, that \(\left\langle
V_{20}\right\rangle =f_{20}\left\langle v_{20}\right\rangle \left(\sigma _3\right)\)
does not depend on \(\sigma _1\) and is free of the angles \(l_5,l_6\) up to
\(O(e^5)\) (i.e. \(\left\langle u_{20}\right\rangle=0\)).

\subsection*{The term \(\left\langle V_{22}\right\rangle\):}

\np The expression proportional to \(C_{22}\) in \(V_G\)
is of the form:

\bnq
\frac{R^2}{r^3}\mathfrak{C}_{22}=
3\frac{R^2}{r^3}\left(\hat{x}^2-\hat{y}^2\right) \ .
\enq

\np The averaged term up to $O(e^5)$ becomes:

\bnq
\left\langle V_{22}\right\rangle =-\frac{R^2}{256a^3}
\left(\left\langle v_{22}\right\rangle \left(\sigma _1,\sigma _3\right)+
\left\langle u_{22}\right\rangle\left(\sigma _1,\sigma _3,l_5\right)\right) + O(e^5) \ ,
\enq

\np with
\bna
&&\left\langle v_{22}\right\rangle =E_1\left([1]_{22}\cos \left(2\sigma _1\right)+
[2]_{22}\cos \left(2\sigma _1+\sigma _3\right)+\right. \\
&&\left.[3]_{22}\cos \left(2\sigma _1+2\sigma _3\right)+
[4]_{22}\cos \left(2\sigma _1+3\sigma _3\right)+
[5]_{22}\cos \left(2\sigma _1+4\sigma _3\right)\right) \ ,
\ena

\np where we collect the terms depending on the orbital eccentricity $e$ by:
\bnq
E_1=e\left(-56+123 e^2\right) \ .
\enq

\np The time dependent part of the expression is

\bna
&&\left\langle u_{22}\right\rangle =318e^3\left([6]_{22}\cos \left(2\sigma _1+2l_5\right)+
[7]_{22}\cos \left(2\sigma _1+\sigma _3+2l_5\right)+\right. \\
&&\left.[8]_{22}\cos \left(2\sigma _1+2\sigma _3+2l_5\right)+
[9]_{22}\cos \left(2\sigma _1+3\sigma _3+2l_5\right)+
[10]_{22}\cos \left(2\sigma _1+4\sigma _3+2l_5\right)\right) \ ,
\ena

\np where the coefficients $[1]_{22}\dots[10]_{22}$, depending on the resonant
actions, are given in the  Appendix. Note, that \(\left\langle v_{22}\right\rangle\)
enters a term with the only resonant argument \(2\sigma _1\), while the other
Fourier modes are of the form
\bnq
2\sigma _1+k \sigma _3, k=1,2,3,\ldots
\enq

\np up to order \(O(e^4)\). The terms \(\left\langle u_{22}\right\rangle
=\left\langle u_{22}\right\rangle \left(\sigma _1,\sigma _3,l_5\right)\) do not
depend on \(l_6\) and contain terms of the form 
\bnq
2\sigma _1+k \sigma _3+2l_5, k=1,2,3,\ldots
\enq

\np with the additional argument \(2l_5\).

\subsection*{The term \(\left\langle V_{30}\right\rangle\):}

\np The third order term proportional to \(C_{30}\) turns out to be:

\bnq{}
\frac{R^3}{r^4}\mathfrak{C}_{30}=
\frac{1}{2}\frac{R^3}{r^4}\hat{z}\left(5\hat{z}^2-3\right).
\enq

\np After the averaging no pure resonant terms survive (\(\left\langle
	v_{30}\right\rangle=0\)) and \(\left\langle V_{30}\right\rangle\) takes the form

\bnq{}
\left\langle V_{30}\right\rangle =\frac{R^3}{128a^4}\left\langle
u_{30}\right\rangle \left(\sigma _3,l_5\right)
\enq

\np with
\bna
&&\left\langle u_{30}\right\rangle =E_2\left([1]_{30}\sin \left(l_5-3\sigma _3\right)+
[2]_{30}\sin \left(l_5-2\sigma _3\right)+\right. \\
&&[3]_{30}\sin \left(l_5-\sigma _3\right)+[4]_{30}\sin \left(l_5\right)+
[5]_{30}\sin \left(l_5+\sigma _3\right)+\\
&&\left.[6]_{30}\sin \left(l_5+2\sigma _3\right)+[7]_{30}\sin \left(l_5+3\sigma _3\right)\right).
\ena

\np In the above expression the $[j]_{30}$, with $j=1,\dots7$, label the third order 
coefficients, depending on the resonant actions, which are collected together in the
Appendix. Moreover, we define

\bnq
E_2=e \left(2+5 e^2\right) \ 
\enq

\np to collect the contributions, which depend on the eccentricity. Note, that the
time dependent terms \(\left\langle u_{30}\right\rangle\) are again independent of
\(l_6\) (they only depend implicitly on it through the definition of \(\sigma
_1,\sigma _3\)). We also conclude, that $C_{30}$ does not contribute to the
time independent resonant model, at the first order of masses approximation.

\subsection*{The term \(\left\langle V_{40}\right\rangle\):}

\np The expression proportional to \(C_{40}\) of \(\left\langle
V_G\right\rangle _{\sigma _4}\) originates from the term

\bnq
\frac{R^4}{r^5}\mathfrak{C}_{40}=\frac{1}{8}\frac{R^4}{r^5}
\left(3-30\hat{z}^2+35\hat{z}^4\right) \ ,
\enq

\np and can again be split into

\bnq
\left\langle V_{40}\right\rangle =\frac{R^4}{4096a^5}\left(\left\langle v_{40}\right\rangle
\left(\sigma _3\right)+\left\langle u_{40}\right\rangle
\left(\sigma _3,l_5\right)\right) \ .
\enq

\np Here the time-independent part is given by the expression:

\bnq
\left\langle v_{40}\right\rangle =E_3\left([1]_{40}+[2]_{40}\cos \left(\sigma _3\right)+
[3]_{40}\cos \left(2\sigma _3\right)+[4]_{40}\cos \left(3\sigma _3\right)+[5]_{40}\cos
\left(4\sigma _3\right)\right) \ ,
\enq

\np where
\bnq
E_3=\left(8+40e^2+105e^4\right) \ ,
\enq

\np and the time dependent contributions are:

\bna
&&\left\langle u_{40}\right\rangle =E_4\left([6]_{40} \cos \left(2 l_5\right)+
[7]_{40}\cos \left(2 l_5-4 \sigma _3\right)+
[8]_{40} \cos \left(2 l_5-3 \sigma _3\right)+\right. \\
&&[9]_{40} \cos \left(2 l_5-2 \sigma _3\right)+
[10]_{40} \cos \left(2 l_5-\sigma _3\right)+
[11]_{40} \cos \left(2 l_5+\sigma _3\right)+ \\
&&\left.[12]_{40} \cos \left(2 l_5+2 \sigma _3\right)+[13]_{40} \cos
\left(2 l_5+3 \sigma _3\right)+[14]_{40} \cos \left(2 l_5+4 \sigma _3\right)\right) \ ,
\ena

\np where
\bnq
E_4=e^2 \left(2+7 e^2\right)
\enq

\np (the coefficients $[\dots]_{40}$ can be found again in the Appendix). Note, that
\(\left\langle v_{40}\right\rangle\) just depends on \(\sigma _3\),
while \(\left\langle u_{40}\right\rangle\) depends on \(\sigma _3,2l_5\) but
not on \(l_6\).

\np To summarize, we find:

\begin{itemize} 

\item[(i)] The effect proportional to $C_{20}$ (or $J_2$) is of order $R^2/a^3$,
is time independent, and depending on the resonant angle $\sigma_3$ only.

\item[(ii)] The effect proportional to $C_{22}$ can be split into time dependent as
well as time independent terms: the time independent terms are proportional to
$e R^2/a^3$, while the time dependent contributions are of order $e^3R^2/a^3$. The former
contains a Fourier term depending just on the resonant argument $2\sigma_1$, while the latter
is depending on integer combinations of $\sigma_1$, $\sigma_3$ and $l_5$ only.

\item[(iii)] There is no time independent effect proportional to $C_{30}$ (or $J_3$) on
the averaged dynamics. Only time dependent terms, which are of order $e R^3/a^4$,
and depending on integer combinations of $\sigma_3$ and $l_5$ contribute to it.

\item[(iv)] There is an important, time independent contribution, of order $R^4/a^5$,
which is proportional to $C_{40}$ and just depending on the resonant angle $\sigma_3$.
The time dependent part of the potential is of order $e^2 R^4/a^5$ only.

\end{itemize}
 
\np The preceeding list shows, that the time dependent effects are smaller than
the time independent. For the numerical integration of the averaged system 
(see Sec.\ref{NUMS}) we are going to use the full potential of the form \equ{EqVt},
while for the analytical study we are going to use the time independent part of
the potential only.

\section{Analytical treatment of the Cassini State 1\label{sec:analytical}}

\np One of our goals is to find a formula similar to Peale's (Eq.\ref{eq:peale}). For that, we start from our averaged Hamiltonian and make the assumption that the orbital quantities are constant. These quantities are the mean motion $n$, the eccentricity $e$, the inclination $i$, and the regression rate of the ascending node $\dot{\ascnode}$. Here the quantities $i$ and $\ascnode$ are defined with respect to a Laplace Plane, that minimizes the variation of the orbital inclination $i$. This means in particular that $i$ and $\ascnode$ are different from the ones given by ephemerides, usually using the ecliptic at J2000.0.

\np To obtain a simple analytical formula giving the obliquity, we neglect the resonant terms in \(\langle V\rangle\), which depend on time through \(l_5=\omega(t),l_6=\ascnode (t)\), and investigate the long-term dynamics close to \(\sigma_1=\sigma _3=0\). The potential \(\langle V\rangle =\langle V\rangle
\left(\sigma _1,\sigma _3\right)\) reduces to

\bnq
\langle V\rangle \left(\sigma _1,\sigma _3\right)=
C_{20}f_{20}\left\langle v_{20}\right\rangle\left(\sigma _3\right)+
C_{22}f_{22}\left\langle v_{22}\right\rangle\left(\sigma _1,\sigma _3\right)+
C_{40}f_{40}\left\langle v_{40}\right\rangle\left(\sigma _3\right) \ ,
\enq

\np the integrable parts reduce to

\bnq
h_0=\mathcal{H}_K+\mathcal{H}_R - \Sigma_1\dot\omega +
\left(\Sigma _3-\Sigma _1\right)\dot{\ascnode} \ ,
\enq

\np and the new Hamiltonian model becomes:

\beq{EqM0}
\mathcal{H}_{I0}=
h_0
-G_c M m
\bigg(
C_{20}f_{20}\left\langle v_{20}\right\rangle +
C_{22}f_{22}\left\langle v_{22}\right\rangle +
C_{40}f_{40}\left\langle v_{40}\right\rangle
\bigg) \ .
\eeq

\np The requirement to remain at the equilibrium is that \(\dot{\Sigma
}_1=\dot{\Sigma }_3=0\) and translates into the set of equations:

\bea{EqF0}
&&f_1\left(\Sigma _1,\Sigma _3\right)\equiv \left(\frac{\partial \mathcal{H}_{I0}}
{\partial \Sigma _1}\right)_{\sigma _1,\sigma _3=0}=0 \ , \nonumber \\
&&f_2\left(\Sigma _1,\Sigma _3\right)=\left(\frac{\partial \mathcal{H}_{I0}}
{\partial \Sigma _3}\right)_{\sigma _1,\sigma _3=0}=0 \ .
\eea

\np The system can be solved for \(\Sigma _1=\Sigma _{1*}\), and \(\Sigma _3=\Sigma
_{3*}\), which implies the {`}equilibrium norm{'} of the angular momentum
\(G_*\) and the {`}equilibrium obliquity{'} \(K_*\), which comes from the
equations \(\Sigma _{1*}=G_*\), \(\Sigma _{3*}=\Sigma _{1*}\left(1-\cos
\left(K_*\right)\right)\). The physical interpretation of the equilibrium
solution is the following: \(\sigma _1=0\) means that the axis of smallest
inertia points (on average) towards the Sun, \(\sigma _3=0\) ensures that the
node of the equator of Mercury is locked with the node of its orbit. While
\(G_*\) defines a small correction of the unperturbed spin frequency, the angle
\(K_*\) defines a specific value of the inertial obliquity.  It translates to
the usual called obliquity $\epsilon$ by the relation

\beq{EqAux}
\cos (\epsilon )=\cos (i)\cos (K)+\sin (i)\sin (K)\cos
\left(\sigma _3\right) \ ,
\eeq

\np which reduces to

\bnq
\epsilon _*=i-K_*
\enq

\np for \(\sigma _3=0\). This corresponds to the 
third Cassini Law \citep{c1693,c1966,n2009} stating that the Laplace normal, 
spin and orbit normal are coplanar.

\np The contributions \(\left\langle v_{nm}\right\rangle\) depend on \(\left(\Sigma
_1,\Sigma _3\right)\) through \( \sin K = s_K=s_K\left(\Sigma _1,\Sigma
_3\right), \cos K = c_K=c_K\left(\Sigma _1,\Sigma _3\right)\), since from \(\Sigma
_1=L_1\), \(\Sigma _3=L_3\) we find \(\Sigma _1=G\), \(\Sigma _3=\Sigma
_1(1-\cos (K))\). We aim to express \equ{EqF0} in terms of \((G,K)\) and thus have
also to express the derivatives \(\partial \left/\partial \Sigma _1\right.\),
\(\partial \left/\partial \Sigma _3\right.\) in terms of \((G,K)\) too. A
simple calculation shows for \(s_K\) and \(c_K\) and its derivatives:

\bnq
\frac{\partial c_K}{\partial \Sigma _1}=\frac{1-c_K}{G} \ , \
\frac{\partial c_K}{\partial \Sigma _3}=-\frac{1}{G} \ , \
\frac{\partial s_K}{\partial \Sigma_1}=\frac{c_K-1}{G t_K} \ , \
\frac{\partial s_K}{\partial \Sigma _3}=\frac{1}{G t_K} \ ,
\enq

\np where we have introduced \(t_K=\tan (K)\), and used the relation
\(s_K{}^2+c_K{}^2=1\) and \(0\leq K\leq \pi /2\) to simplify the expressions. A
long but straightforward calculation shows that \equ{EqF0} can be written as:
\bea{EqF1}
&&f_1(G,K)=-\frac{3 n}{2}+\frac{G-C \left(\dot{\omega}+\dot{\ascnode }\right)}{C}+
\frac{G_C M m}{G}
\left(\frac{R^2}{a^3}\left([1]_f C_{20}+[2]_f C_{22}\right)+
\frac{R^4}{a^5}[3]_f C_{40}\right)=0 \ , \nonumber \\
&&f_2(G,K)=\dot{\ascnode }+\frac{G_C M m}{G} \left(\frac{R^2}{a^3}\left([4]_f C_{20}+
[5]_f C_{22}\right)+\frac{R^4}{a^5}[6]_f C_{40}\right) =0 \ ,
\eea

\np with
\bna
&&[1]_f=-\frac{3}{32} \left(8+12 e^2+15 e^4\right) s_{2 (i-K)} t_{K/2} \ , \\
&&[2]_f=-\frac{3}{8} e \left(-56+123 e^2\right) c_{(i-K)/2}^3 s_{(i-K)/2} t_{K/2} \ , \\
&&[3]_f=\frac{15}{1024} \left(8+40 e^2+105 e^4\right)
\left(2 s_{2 (i-K)}+7 s_{4 (i-K)}\right) t_{K/2}
\ena

\np and
\bna
&&[4]_f=\frac{3}{32}\left(8+12e^2+15e^4\right)s_{2(i-K)}/s_K
=-[1]_f/(2s_{K/2}^2)  \ , \\
&&[5]_f=\frac{3}{64}e\left(-56+123e^2\right)
s_{i-K}^3/s_{(i-K)/2}^2/s_K
=-[2]_f/(2s_{K/2}^2) \ , \\
&&[6]_f=-\frac{15}{1024}\left(8+40e^2+105e^4\right)
\left(2s_{2(i-K)}+7s_{4(i-K)}\right)/s_K
=-[3]_f/(2s_{K/2}^2)\ . 
\ena

\np Note, that we can use the second part of \equ{EqF1} to eliminate \(G\) from the
remaining equation to get:
\beq{EqF2}
F(K)=-\left(\frac{3 n}{2}+\dot{\omega}+\dot{\ascnode } c_K\right)+
G_CM m\left(\frac{R^2}{a^3}
\left([1]_F C_{20}+[2]_F C_{22}\right)+\frac{R^4}{a^5}[3]_F C_{40}\right)=0 \ ,
\eeq

\np with
\bna
&&[1]_F=-\frac{[4]_f}{C\dot\Omega}=-\frac{3}{32C\dot{\ascnode}s_K}
\left(8+12e^2+15e^4\right)s_{2(i-K)} \ , \\
&&[2]_F=-\frac{[5]_f}{C\dot\Omega}=-\frac{3}{64C \dot{\ascnode}
s_Ks_{(i-K)/2}{}^2}e\left(-56+123e^2\right)s_{i-K}{}^3 \ , \\
&&[3]_F=-\frac{[6]_f}{C\dot\Omega}=\frac{15}{1024C \dot{\ascnode }s_K}
\left(8+40e^2+105e^4\right)
\left(2s_{2(i-K)}+7s_{4(i-K)}\right) \ .
\ena

\np The relation Eq.~\ref{EqF2} can be solved for the angle $K$ in an implicit
way. We now substitute \equ{EqAux} {for \(\sigma _3=0\)
in \equ{EqF2} and expand around \(\epsilon =0\) up to first order. Moreover we
make use of the relations \(C=c m R^2\) and \(n^2a^3\simeq G_C M\) to get rid of
some constants.  Within these approximations we arrive at the simple formula:

\bea{EqF3}
\epsilon &&=\left(1+\frac{2}{3}\frac{\dot{\ascnode }}{n}\cos (i)+\frac{2\dot{\omega}}{3n}\right)  \times \\ 
&&\frac{c \dot{\ascnode } \sin (i)}{n\left(2C_{22}
\left(\frac{7}{2}e-\frac{123}{16}e^3\right)-
C_{20}\left(1+\frac{3}{2}e^2+\frac{15}{8}e^4\right)+
C_{40}\left(\frac{R}{a}\right)^2\left(\frac{5}{2}+
	\frac{25}{2}e^2+\frac{525}{16}e^4\right)-
\frac{2}{3}\left(\frac{\dot{\ascnode}}{n}\right)^2c
\sin (i)^2\right)} \ . \nonumber
\eea

\np Notice, that here the parameter \(\dot{\ascnode }\) is assumed to be
negative. This yields a negative obliquity, in the
following we use a positive value for $\epsilon$, in fact $|\epsilon|$. The
formula coincides for \(C_{40}=0\) and $\dot{\omega}=0$ with Peale{'}s formula up to
\(O(\dot{\ascnode }/n)^2\) with the exception of the term \(c \dot{\ascnode
}\cos (i)\) in the denominator \citep{ym2006} \footnote{For small
\((\dot{\ascnode }/n)\) it is possible to simplify the calculations by setting
$G=n_sC\simeq \frac{3}{2}n C$ in the second part of \equ{EqF1}, expanding up to
\(1\)st order in \(\epsilon\) close to \(\epsilon =0\).  The resulting formula
for \(\epsilon\) coincides with \equ{EqF3} up to \(O(\dot{\ascnode}/n^2)\). The
additional terms are therefore stemming from the small spin frequency
correction, not taken into account in (Peale 1981).}. A comparison of
\equ{EqF3} with \equ{eq:peale} and with Eq.(4) of \cite{p1981} shows that the
difference between those formulas is less than $1$ arcsecond within the interval
$0.3\leq c \leq0.4$, with an offset of about \(10\epsilon ^2\simeq 600mas\) in
the case of Mercury, compared with the formulae given in \citep{ym2006} and
\citep{p1981}. Note that the parameter \(C_{30}\) is absent in this simple
averaged model (there is however a time dependent effect as we will see below),
the influence of \(C_{40}\) on the denominator is of the order of
\(O(R/a)^2\simeq 4.\cdot 10^{-5}\), and \(O(\dot{\ascnode}/n)^2\simeq 8.\cdot
10^{-7}\) does not modify the results. The quantity $2/3\left(\dot{\ascnode}\cos(i)+\dot{\omega}\right)/n\simeq7.\cdot10^{-7}$
has also a negligible influence, this supports the omission of $\dot{\omega}$ in Peale's formula.

\np We conclude the section with a short summary of the assumptions, which were
made to obtain \equ{EqF3}.

\begin{itemize}
 \item[i)] zero wobble,
 \item[ii)] only the coupling of the spin on the orbit was taken into account,
 \item[iii)] short periodic effects are neglected (average over mean orbital
motion and time),
 \item[iv)] the orbital parameters $i$, $e$, $n$, $\dot{\ascnode}$ and $\dot{\omega}$ are kept constant.
\end{itemize}

\np The resulting accuracy of the formula can be quantified as follows (errors
related to angles are given in radians):
\begin{itemize}
\item[1)] for the average a 4th order expansion in eccentricity \(e\) was taken 
into account (error \(O(e^5)\simeq 3.\cdot 10^{-4}\)),
\item[2)] time dependent resonant terms are neglected (with this we set 
\(d \epsilon /d t=0\)),
\item[3)] a first order expansion in obliquity \(\epsilon\) allowed to make the
formula explicit and induced an additional error of 
\(O(\epsilon^2)\simeq1.\cdot 10^{-3}\).
\end{itemize}

\np This formula and Peale's formula as well are lacking of the fact, that
the obliquity changes with time not only because $\dot{\ascnode}$ is not
a constant but because the 'equilibrium' conditions $\sigma_1=\sigma_3=0$
cannot be fulfilled for all times. As we have seen, even for constant
precession rates, the Hamiltonian is not
a conserved quantity, i.e. for a Hamiltonian of the form $\mathcal{H}=\mathcal{H}(\Sigma_1,
\Sigma_3,\sigma_1,\sigma_3,t)$ we cannot deduce that $\sigma_1=\sigma_3=0$
for all times, since

\bnq
\dot\sigma_{1,3}=\frac{\partial \mathcal{H}(\Sigma_1,\Sigma_3,\sigma_1,\sigma_3,t)}
{\partial\Sigma_{1,3}}\to\dot\sigma_{1,3}\neq0 \ ,
\enq

\np and therefore $\sigma_{1,3}$ are subject to change too. A second
(and probably more visible) concern of formulas of the type \equ{EqF3}
is the fact, that the resulting $\epsilon$ strongly depends on
the choice of the reference plane to which the orbital inclination
$i$ and in which the average precession rate $\dot{\ascnode}$ are
defined. To optimize the result, the reference plane should be the
plane to which the variations in $i$ become minimal, which can 
be seen as a generalization to the so-called Laplace plane 
(not to be confused with the invariant Laplace plane, the plane
normal to the angular momentum of the complete system). In the next
Section we present a possible solution to these kinds of problems.

\section{Including the secular variations of the orbital elements\label{NUMS}}

\par Up to now we have considered, as Peale did, that the orbital quantities $n$, $e$, $i$, $\dot{\omega}$  and $\dot{\ascnode}$ were constant. In \citep{nd2012} a numerical method considering the variations of the orbital elements is proposed. We will here use the same method, with the refinements that more spherical harmonics are considered, the second-order spherical harmonics of Mercury $J_2$ and $C_{22}$ are known with a much better accuracy, and the averaged equations are expanded up to a higher degree in eccentricity  / inclination.

\subsection{Influence on one example}

\par In a reference frame based on the ecliptic at J2000, we define averaged eccentricities and inclinations (Eq. 6 to 10 of \citep{nd2012}) as:

\begin{eqnarray}
h(t) & = & -7.76651\times10^{-11}t^2+1.43999\times10^{-6}t+0.200722, \label{eq:h} \\
k(t) & = & -2.31417\times10^{-11}t^2-5.52628\times10^{-6}t+0.0446629, \label{eq:k} \\
p(t) & = & 2.38036\times10^{-16}t^3-9.03918\times10^{-12}t^2-1.27635\times10^{-6}t+0.0456355, \label{eq:p3} \\
p(t) & = & -1.04673\times10^{-11}t^2-1.27792\times10^{-6}t+0.0456362, \label{eq:p2} \\
q(t) & = & 2.52407\times10^{-16}t^3-1.06586\times10^{-11}t^2+6.54322\times10^{-7}t+0.0406156, \label{eq:q3} \\
q(t) & = & -1.21729\times10^{-11}t^2+6.52656\times10^{-7}t+0.0406163, \label{eq:q2}
\end{eqnarray}
with $h(t)=e(t)\sin\varpi(t)$, $k(t)=e(t)\cos\varpi(t)$, $p(t)=\sin\left(\frac{i(t)}{2}\right)\sin\ascnode(t)$ and 
$q(t)=\sin\left(\frac{i(t)}{2}\right)\cos\ascnode(t)$, the time origin begin J2000 and the time unit being the year. 
The inclination of Mercury $i$ is here defined with respect to the ecliptic at J2000, and $\varpi=\omega+\ascnode=l_5+l_6$.
The Eq.\ref{eq:h} to \ref{eq:q2} are fits over the duration of the JPL DE406 ephemerides, i.e. 6,000 years. For the inclination variables, 2 fits 
are given: the third-order fit (Eq.\ref{eq:p3} and Eq.\ref{eq:q3}) is more accurate, but the second-order one is easier 
to extrapolate into a trigonometric decomposition.

\par From these formulae we extract trigonometric expressions (Eq.24 to 27 in \citep{nd2012}):

\begin{eqnarray}
h(t) & = & 0.1990903983\sin\varpi_1(t)+0.01094807206\sin\varpi_2(t), \label{eq:htrig} \\
k(t) & = & 0.1990903983\cos\varpi_1(t)+0.01094807206\cos\varpi_2(t), \label{eq:ktrig} \\
p(t) & = & 0.06094690052\sin\Omega_1(t)+0.01442538649\sin\Omega_2(t), \label{eq:ptrig} \\
q(t) & = & 0.06094690052\cos\Omega_1(t)+0.01442538649\cos\Omega_2(t), \label{eq:qtrig} 
\end{eqnarray}
with

\begin{eqnarray}
\varpi_1(t) & = &  2.852011398\times10^{-5}t+1.30845314198, \label{eq:varpi1} \\
\varpi_2(t) & = &  4.767836272\times10^{-6}t+2.26085090227, \label{eq:varpi2} \\
\Omega_1(t) & = & -2.298222197\times10^{-5}t+0.60658814513, \label{eq:Omega1} \\
\Omega_2(t) & = &  1.340719884\times10^{-5}t+2.28580288184. \label{eq:Omega2}
\end{eqnarray}

\par Using trigonometric series make the orbital solutions easy to extrapolate without divergence. We extrapolate them over several millions of years to optimize their numerical identification, and we use it to perform numerical integrations of the Hamilton equations derived from the averaged Hamiltonian $\mathcal{H}_{I0}$ (Eq.16). We assume that the resulting obliquity is close to the real one over the validity of the DE406 ephemerides. This assumption will be checked in Sect.\ref{sec:verif}.

\par We can now plug these new orbital elements in the equations of the averaged rotational dynamics of Mercury. Using Laskar's frequency analysis \citep{l1999,l2005}, we express the ecliptic obliquity $K$ (the ecliptic at J2000 being our inertial reference plane) and the resonant argument $\sigma_3$ with a quasiperiodic decomposition such as

\begin{eqnarray}
	K(t) & = & i(t)+\sum a_i\cos\left(\omega_i t+\phi_i\right), \label{eq:Knaff} \\
	\sigma_3(t) & = & \sum b_j\cos\left(\omega_j t+\phi_j\right), \label{eq:sig3naff}
\end{eqnarray}
$a_i$ and $b_j$ being real amplitudes, $\omega_{i,j}$ frequencies, and $\phi_{i,j}$ phases at $t=0$. The frequencies can either come from the forced motion, and so are combinations of the ones present in the orbital motion (Eq.\ref{eq:varpi1} to Eq.\ref{eq:Omega2}), or are due to free librations, that are expected to be damped. Their presence in the numerical outputs comes from a non optimal choice of the initial conditions. Our first run with $C_{20}=-5.031\times10^{-5}$, $C_{22}=8\times10^{-6}$, $C_{30}=-1.188\times10^{-5}$, $C_{40}=-1.95\times10^{-5}$ and $C=0.35mR^2$ yields the Tab.\ref{tab:Ksig1}.

\begin{table}[ht]
	\centering
	\caption{An example of frequency analysis of the numerical outputs of our system.\label{tab:Ksig1}}
	\begin{tabular}{rrrr}
		\hline
		$N$ & Amplitude & Period & Identification \\
		& (arcmin) & (y) & \\
		\hline
		$K-i$ & & & \\
		\hline
		$1$ & $1.9755$ &  $\infty$ & $<K-i>$ \\
		$2$ & $0.2682$ & 172,665.2 & $\Omega_2-\Omega_1$ \\
		$3$ & $0.0592$ &  86,332.6 & $2\Omega_2-2\Omega_1$ \\
		$4$ & $0.0254$ & 264,530.5 & $\omega_1-\omega_2$ \\
		$5$ & $0.0132$ &  60,999.1 & $2\omega_1-2\Omega_1$ \\
		$6$ & $0.0121$ &  57,555.3 & $3\Omega_2-3\Omega_1$ \\
		$7$ & $0.0026$ &   4,167.1 & free \\
		$8$ & $0.0025$ &  43,166.2 & $4\Omega_2-4\Omega_1$ \\
		$9$ & $0.0025$ &   4,302.6 & free \\
		\hline\hline
		$\sigma_3$ & & & \\
		\hline
		  $1$ & $6.2509$ & 172,665.2 & $\Omega_2-\Omega_1$ \\
		  $2$ & $1.4683$ &  86,332.6 & $2\Omega_2-2\Omega_1$ \\
		  $3$ & $0.3462$ &  57,555.1 & $3\Omega_2-3\Omega_1$ \\
		  $4$ & $0.1112$ &  60,999.0 & $2\omega_1-2\Omega_1$ \\
		  $5$ & $0.0816$ &  43,166.3 & $4\Omega_2-4\Omega_1$ \\
		  $6$ & $0.0400$ & 497,291.8 & $\varpi_2-\varpi_1+\Omega_2-\Omega_1$ \\
		  $7$ & $0.0396$ & 104,473.0 & $\varpi_1-\varpi_2+\Omega_2-\Omega_1$ \\
		  $8$ & $0.0261$ &  45,075.0 & $2\varpi_1-3\Omega_1+\Omega_2$ \\
		  $9$ & $0.0214$ &   4,167.1 & free \\
		 $10$ & $0.0209$ &   4,302.6 & free \\
		 \hline 
 	\end{tabular}
\end{table}

\par The frequency analysis giving this table has been performed over $7.95$ Myr with 4,096 points equally spaced by 1,942.5 years. The identification
of the oscillating arguments has been made in comparing the frequencies and initial phases of these arguments with integer combinations of the proper 
modes (Eq.\ref{eq:varpi1} to \ref{eq:Omega2}). The phases are useful to discriminate between sines and cosines in the decompositions of $K(t)$ and $\sigma_3(t)$. 
The free librations should have periods of $\approx15$ years in longitude\footnote{The period of the expected longitudinal librations is in fact close to 12 years because of the molten outer core. 
In this averaged model we consider that Mercury is a rigid, homogeneous body since it is appropriate to estimate the obliquity.} and $\approx1,000$ years in obliquity \citep{dl2004,rb2004}, these periods should 
appear aliased, while the forced perturbations should not since their periods should be bigger than 10,000 years. We then optimize the initial
conditions iteratively in removing the free librations, as described in \citep{ndc2013}. These free 
oscillations act as a noise in the determination of the forced ones, that explains for instance small discrepancies in the periods. 
Refining these initial conditions improves the accuracy of the determination of the forced oscillations.

\subsection{Fitting the initial conditions}

\par The goal is to find relevant initial conditions for any set of interior parameters $(C_{20},C_{22},C_{30},C_{40},C)$ consistent with 
the observations and the theory. For that we considered 56 differents cases, in letting the interior parameters vary one by one between the uncertainties 
due to MESSENGER data (Tab.\ref{tab:smith}). The range of variations we considered is:

\begin{itemize}

\item $C_{20}\in[-5.1\times10^{-5};-4.9\times10^{-5}]$ (nominal value: $-5.031\times10^{-5}$),

\item $C_{22}\in[8\times10^{-6};8.2\times10^{-6}]$ (nominal value: $8.088\times10^{-6}$),

\item $C_{30}\in[-1.3\times10^{-5},-1.1\times10^{-5}]$ (nominal value: $-1.188\times10^{-5}$),

\item $C_{40}\in[-2.19\times10^{-5};-1.71\times10^{-5}]$ (nominal value: $-1.95\times10^{-5}$),

\item $C/(mR^2)\in[0.32;0.38]$ (nominal value: $0.35$).

\end{itemize}

\par For each of them, a numerical study has been performed as explained above. After refinement of the initial conditions that gave us
an accurate frequency decomposition of the signals, we identified 34 amplitudes, all given by the frequency analysis. We can now write:

\begin{eqnarray}
K & = & i+a_1-2a_2\cos\left(\Omega_2-\Omega_1\right)+2a_3\cos\left(2\Omega_2-2\Omega_1\right)-2a_4\cos\left(\varpi_1-\varpi_2\right) \nonumber \\ 
 & + & 2a_5\cos\left(2\varpi_1-2\Omega_1\right)-2a_6\cos\left(3\Omega_2-3\Omega_1\right)+2a_7\cos\left(4\Omega_2-4\Omega_1\right)\nonumber \\
 & + & 2a_8\cos\left(\varpi_2-\varpi_1+\Omega_2-\Omega_1\right)+2a_9\cos\left(\varpi_1-\varpi_2+\Omega_2-\Omega_1\right)+2a_{10}\cos\left(\varpi_1+\varpi_2-2\Omega_1\right) \nonumber \\
 & - & 2a_{11}\cos\left(2\varpi_1-3\Omega_1+\Omega_2\right)-2a_{12}\cos\left(5\Omega_2-5\Omega_1\right)+2a_{13}\cos\left(2\varpi_1-2\varpi_2\right) \nonumber \\
 & - & 2a_{14}\cos\left(\varpi_1-\varpi_2-2\Omega_1+2\Omega_2\right)-2a_{15}\cos\left(\varpi_2-\varpi_1-2\Omega_1+2\Omega_2\right) \nonumber \\
 & - & 2a_{16}\cos\left(2\varpi_1-2\Omega_2\right), \label{eq:KmI} \\
\sigma_3 & = & 2a_{17}\sin\left(\Omega_2-\Omega_1\right)-2a_{18}\sin\left(2\Omega_2-2\Omega_1\right)+2a_{19}\sin\left(3\Omega_2-3\Omega_1\right) \nonumber \\
 & - & 2a_{20}\sin\left(2\varpi_1-2\Omega_1\right)-2a_{21}\sin\left(4\Omega_2-4\Omega_1\right)-2a_{22}\sin\left(\varpi_2-\varpi_1+\Omega_2-\Omega_1\right) \nonumber \\
 & - & 2a_{23}\sin\left(\varpi_1-\varpi_2+\Omega_2-\Omega_1\right)+2a_{24}\sin\left(2\varpi_1-3\Omega_1+\Omega_2\right)+2a_{25}\sin\left(5\Omega_2-5\Omega_1\right) \nonumber \\
 & + & 2a_{26}\sin\left(2\varpi_1-\Omega_1-\Omega_2\right)-2a_{27}\sin\left(\varpi_1+\varpi_2-2\Omega_1\right)+2a_{28}\sin\left(-\varpi_1+\varpi_2-2\Omega_1+2\Omega_2\right) \nonumber \\
 & + & 2a_{29}\sin\left(\varpi_1-\varpi_2-2\Omega_1+2\Omega_2\right)-2a_{30}\sin\left(2\varpi_1-4\Omega_1+2\Omega_2\right)-2a_{31}\sin\left(6\Omega_2-6\Omega_1\right) \nonumber \\
 & + & 2a_{32}\sin\left(\varpi_1+\varpi_2-3\Omega_1+\Omega_2\right)-2a_{33}\sin\left(\varpi_1-\varpi_2-3\Omega_1+3\Omega_2\right) \nonumber \\
 & - & 2a_{34}\sin\left(\varpi_2-\varpi_1-3\Omega_1+3\Omega_2\right), \label{eq:sig3}
\end{eqnarray}
with

\begin{equation}
\label{eq:ai1}
a_i  =  \frac{C/\left(mR^2\right)}{\alpha_i C/\left(mR^2\right)+\beta_i C_{20}+\gamma_i C_{22}+\delta_i}
\end{equation}
for $i=1,2,5,6,17,18,19,20,21,22,23,24,26,27$,

\begin{equation}
\label{eq:ai2}
a_i  =  \frac{C/\left(mR^2\right)}{\alpha_i+\beta_i C_{20}+\gamma_i C_{22}}
\end{equation}
for $i=3,4$

\begin{equation}
\label{eq:ai3}
a_i  =  \frac{C/\left(mR^2\right)}{\alpha_i+\beta_i C_{20}}
\end{equation}
for $i=7,8,9,10,11,12,13,14,15,16,34$, and

\begin{equation}
\label{eq:ai4}
a_i  =  \frac{C/\left(mR^2\right)}{\alpha_i C/\left(mR^2\right)+\beta_i C_{20}+\gamma_i}
\end{equation}
for $i=25,28,29,30,31,32,33$.

\par The form of these formulae comes from Peale's formula (Eq.\ref{eq:peale} \& Eq.\ref{EqF3}). We expected to get amplitudes alike

\begin{equation}
	\label{eq:hope}
	a_i=\frac{C/(mR^2)}{\alpha_i C/(mR^2)+\beta_i C_{20}+\gamma_i C_{22}+\zeta_i C_{30}+\phi_i C_{40}+\eta_i}.
\end{equation}
For that, we tried to fit amplitudes with respect to one parameter, i.e.

\begin{equation}
\label{eq:CMR21}
f\left(\frac{C}{mR^2}\right)=\frac{aC/\left(mR^2\right)}{1+bC/\left(mR^2\right)}
\end{equation}
when possible and

\begin{equation}
\label{eq:CMR22}
f\left(\frac{C}{mR^2}\right)=aC/\left(mR^2\right)
\end{equation}
when not, and also

\begin{equation}
\label{eq:C20}
f\left(C_{20}\right)=\frac{1}{a+bC_{20}},
\end{equation}

\begin{equation}
\label{eq:C22}
f\left(C_{22}\right)=\frac{1}{a+bC_{22}},
\end{equation}

\begin{equation}
\label{eq:C30}
f\left(C_{30}\right)=\frac{1}{a+bC_{30}},
\end{equation}

\begin{equation}
\label{eq:C40}
f\left(C_{40}\right)=\frac{1}{a+bC_{40}}.
\end{equation}

\par When 2 numbers are present ($a$ and $b$), they are fitted simultaneously. It turned out that it was 
impossible to estimate the influence of $C_{30}$ and $C_{40}$ with
enough reliability, that is the reason why they do not appear in the final formulae (Eq.\ref{eq:ai1} to \ref{eq:ai4}).
The coefficients used are gathered in Tab.\ref{tab:coeff}.

\begin{table}[ht]
\centering
\caption{Coefficients involved in the Eq.\ref{eq:ai1} to \ref{eq:ai4}.\label{tab:coeff}}
\begin{tabular}{r|rrrr}
i & $\alpha_i$ & $\beta_i$ & $\gamma_i$ & $\delta_i$ \\
\hline
 1 & $-9.6916394157$ &     $-1.022791\times10^7$ &      $1.224118\times10^7$ &  $-0.041284174514$ \\
 2 &  $57.319667133$ &     $-1.495053\times10^8$ &     $1.8067315\times10^8$ & $-15.290923597$ \\
 3 &   $-288.588048$ &      $-6.75794\times10^8$ &      $8.648745\times10^8$ & --  \\
 4 &   $109.5839605$ &    $-2.3327045\times10^9$ &    $-2.8315945\times10^9$ & -- \\
 5 &  $9618.2490875$ &     $-5.967955\times10^9$ & $-1.4620515\times10^{10}$ & $-4270.5575456$ \\
 6 &  $5690.3083952$ &      $-3.29021\times10^9$ &       $4.59872\times10^9$ & $-5791.2841683$ \\
 7 &     $160882.75$ &  $-1.563842\times10^{10}$ &  --                       & -- \\
 8 &    $-330146.25$ & $-3.3956545\times10^{10}$ &  --                       & -- \\
 9 &    $-307545.35$ &  $-3.441592\times10^{10}$ &  --                       & -- \\
10 &     $-879441.5$ &   $-4.59746\times10^{10}$ &  --                       & -- \\
11 &    $-1025209.5$ &   $-5.12722\times10^{10}$ &  --                       & -- \\
12 &        $760424$ &  $-7.318535\times10^{10}$ &  --                       & -- \\
13 &      $-3583265$ & $-1.7097535\times10^{11}$ &  --                       & -- \\
14 &      $-1352596$ &  $-1.522164\times10^{11}$ &  --                       & -- \\
15 &    $-1475953.5$ &  $-1.554497\times10^{11}$ &  --                       & -- \\
16 &      $-4788455$ & $-2.4530905\times10^{11}$ &  --                       & -- \\
17 &  $0.0459703076$ &               $-111936.3$ &               $133840.35$ & $0.00068014043987$ \\
18 &  $0.5097512707$ &                 $-474698$ &                  $568708$ & $-0.0063487867442$ \\
19 &  $3.6758306831$ &                $-2005542$ &                 $2431534$ & $-0.32882074509$ \\
20 &  $20.470309592$ &               $-12322835$ &               $-31382330$ & $1.6246447377$ \\
21 &  $22.351343806$ &                $-8470000$ &                $10602865$ & $-4.3502813922$ \\
22 &  $1.8968370715$ &               $-25710160$ &               $-28807205$ & $-14.991580755$ \\
23 &  $28.007891612$ &               $-25704595$ &               $-30518425$ & $0.75276141087$ \\
24 &  $151.69560968$ &               $-52410050$ &              $-127924300$ & $-54.254151937$ \\
25 &  $11.320467298$ &               $-35729050$ &            $384.87733645$ & -- \\
26 &  $60.949766585$ &               $-91723450$ &              $-262171000$ & $275.67522095$ \\
27 &  $93.149257619$ &               $-96422550$ &              $-265414800$ & $257.09346578$ \\
28 &  $256.01109782$ &              $-108893750$ &           $-1098.8078842$ & -- \\
29 &  $93.896968203$ &              $-109386550$ &           $-1017.8654389$ & -- \\
30 &  $880.39601125$ &              $-223772150$ &           $-4722.8636039$ & -- \\
31 & $-390.02956574$ &              $-145710250$ &            $2105.3653480$ & -- \\
32 &  $1929.3930214$ &              $-404092500$ &           $-8327.8625575$ & -- \\
33 &  $2613.8902647$ &              $-456302000$ &           $-4855.3715926$ & -- \\
34 &    $-2905.1715$ &              $-438245500$ &  --                       & --
\end{tabular}
\end{table}

\section{The influence of the different effects\label{sec:influence}}

\par The derivation of these new formulae for the obliquity of Mercury allows us to estimate the influence of usually neglected effects,
like the secular variations of the orbital elements, the tides and the higher order harmonics. For that we first need to test
the reliability of our initial conditions on a real, non-averaged simulation of the rotation of Mercury.

\subsection{Differences between the averaged and the unaveraged system\label{sec:verif}}

\par We proceed as in \citep{nd2012}, Sect.4. To simulate the non-averaged rotation of Mercury, we integrate numerically the equations 
related to the following Hamiltonian:

\begin{eqnarray}
	\mathcal{H} & = & \frac{n}{2(1-\delta)}(P^2-3\delta P) \label{eq:hamil} \\
	& - & \frac{3}{2}\frac{G_CM}{nd^3}\left(\epsilon_1\left(x^2+y^2\right)+\epsilon_2\left(x^2-y^2\right)+\epsilon_3\left(\frac{R}{d}\right)z(5z^2-3)+\epsilon_4\left(\frac{R}{d}\right)^2(3-30z^2+35z^4)\right), \nonumber
\end{eqnarray}
with $\epsilon_1=-C_{20}mR^2/C$, $\epsilon_2=2C_{22}mR^2/C$, $\epsilon_3=C_{30}mR^2/(2C)$, $\epsilon_4=C_{40}mR^2/(8C)$, $\delta=1-C_m/C$, $C_m$ being 
the polar inertial momentum of the mantle of Mercury, P 
is the norm of the angular momentum normalized by $nC$, $d$ is the Sun-Mercury distance, and $x$ and $y$ are the first two coordinates
of the unit vector pointing to the Sun in a reference frame defined by the principal axes of inertia of Mercury. The canonical variables associated
with this Hamiltonian are

\begin{center}
$\begin{array}{lll}
l_1, & \hspace{3cm} & P=\frac{L_1}{nC}, \\
l_3, & \hspace{3cm} & R=\frac{L_3}{nC}, 
\end{array}$
\end{center}
the Hamilton equations associated being

\begin{center}
$\begin{array}{lll}
\frac{dl_1}{dt}=\frac{\partial\mathcal{H}}{\partial P}, & \hspace{3cm} & \frac{dP}{dt}=-\frac{\partial\mathcal{H}}{\partial l_1}, \\
\frac{dl_3}{dt}=\frac{\partial\mathcal{H}}{\partial R}, & \hspace{3cm} & \frac{dR}{dt}=-\frac{\partial\mathcal{H}}{\partial l_3}.
\end{array}$
\end{center}

\par The position of the Sun with respect to Mercury is computed using JPL DE406 ephemerides, so it contains every perturbation, including
the planetary ones. Inappropriate initial conditions in the numerical integration of the equations would yield
unexpected free oscillations. The free longitudinal oscillations, their period being $\approx12$ years, can be easily damped adiabatically over 
5,000 years, the DE406 ephemerides starting at -3,000. However, the free oscillations in obliquity, whose period is about 1,000 years, cannot be
damped over such a timescale without a significant and artificial impact on the equilibrium. So, we add a damping only on the longitudinal motion,
and we get free oscillations in obliquity as in Fig.\ref{fig:obliq}. This obliquity is the actual obliquity $\epsilon$, it is equal to $i-K$
only if $\sigma_3=0$. We obtain it with

\begin{equation}
	\label{eq:epsilon}
	\cos\epsilon=\frac{\vec{G}\cdot\vec{n}}{||\vec{G}||},
\end{equation}
where $\vec{n}$ is the instantaneous normal to the orbit.

\begin{figure}[ht]
\centering
\includegraphics{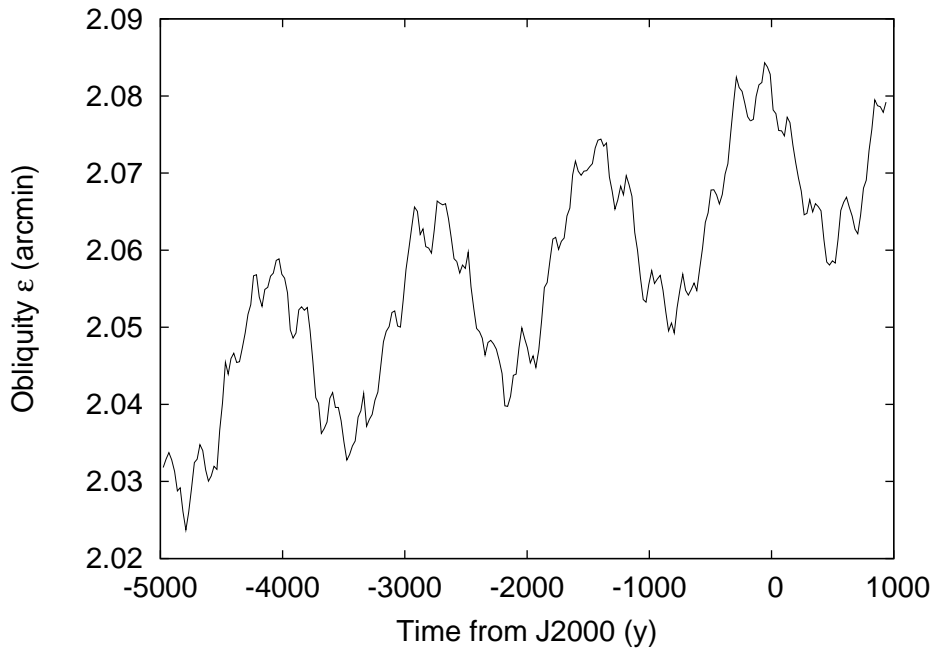}
\caption{Obliquity $\epsilon$ of Mercury given by the unaveraged system, with $C_{20}=-5.031\times10^{-5}$, $C_{22}=8.088\times10^{-6}$, $C_{30}=C_{40}=0$
and $C=0.35mR^2$. The $\approx$ 1,000 years oscillations are free librations that would not be present if the initial conditions were ideal.\label{fig:obliq}}
\end{figure}

\par The amplitude of the free oscillations can be seen as an estimation of the error due to the initial conditions. This error can come from all the approximations made in the averaging process, in particular the limitation to a first order averaging, the expansions in eccentricity, or the exclusion of the planetary perturbations.

\par The Fig.\ref{fig:freeeps} shows the amplitude of these oscillations for 2 orbital theories, JPL DE406 and INPOP10a \citep{flkmdgct2011}, and different values of the interior parameters $C_{20}$, $C_{22}$ and $C$, the other ones not affecting our initial conditions. These amplitudes have been obtained thanks to a frequency analysis. We can see that the amplitude of the free oscillations is always smaller than 750 milli-arcsec. For INPOP10a, many points are missing. The reason is that this theory gives the orbital motion of Mercury over 2 kyr, while DE406 gives it over 6 kyr. The free librations that the frequency analysis is expected to detect have a period of the order of 1 kyr, so in some of the numerical simulations, 2 kyr are not long enough to represent 2 free periods. For this reason, they are sometimes not detected. Usually ephemerides are designed to be very accurate over a quite limited timespan, so 2 kyr should be long enough. But the specific case of a planetary obliquity is a long-term dynamics, for that an orbital theory over several thousands of years is required. Anyway, we can see that no free oscillation with an amplitude bigger than 660 milli-arcsec is detected, this is smaller than the worst case with DE406. 

\citet{mpshgjygpc2012} derived from observations a moment of inertia $C=0.346mR^2$, so the amplitude of the free oscillations should be $\approx650$ milli-arcsec with DE406. This is the theoretical error induced by the Eq.\ref{eq:KmI} and Eq.\ref{eq:sig3}.

\begin{figure}[ht]
\centering
\begin{tabular}{cc}
\includegraphics[width=0.45\textwidth]{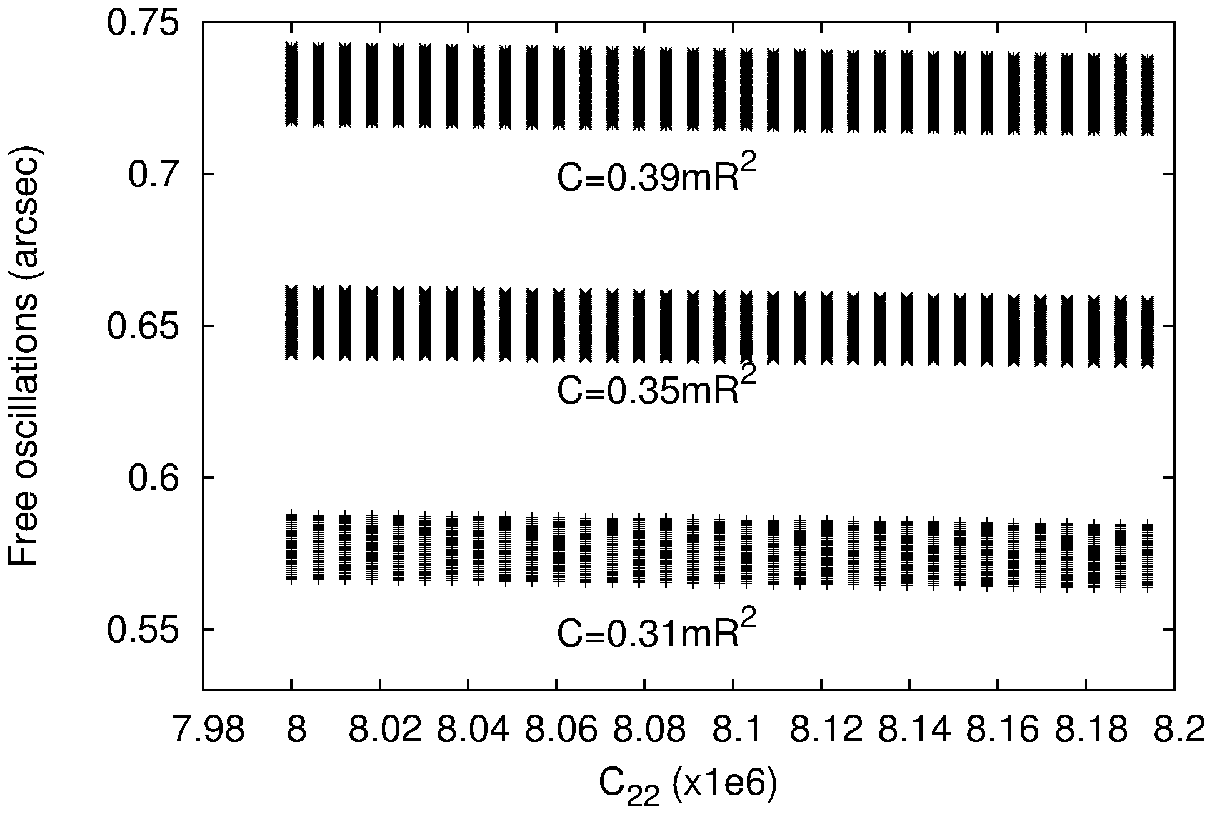} & \includegraphics[width=0.45\textwidth]{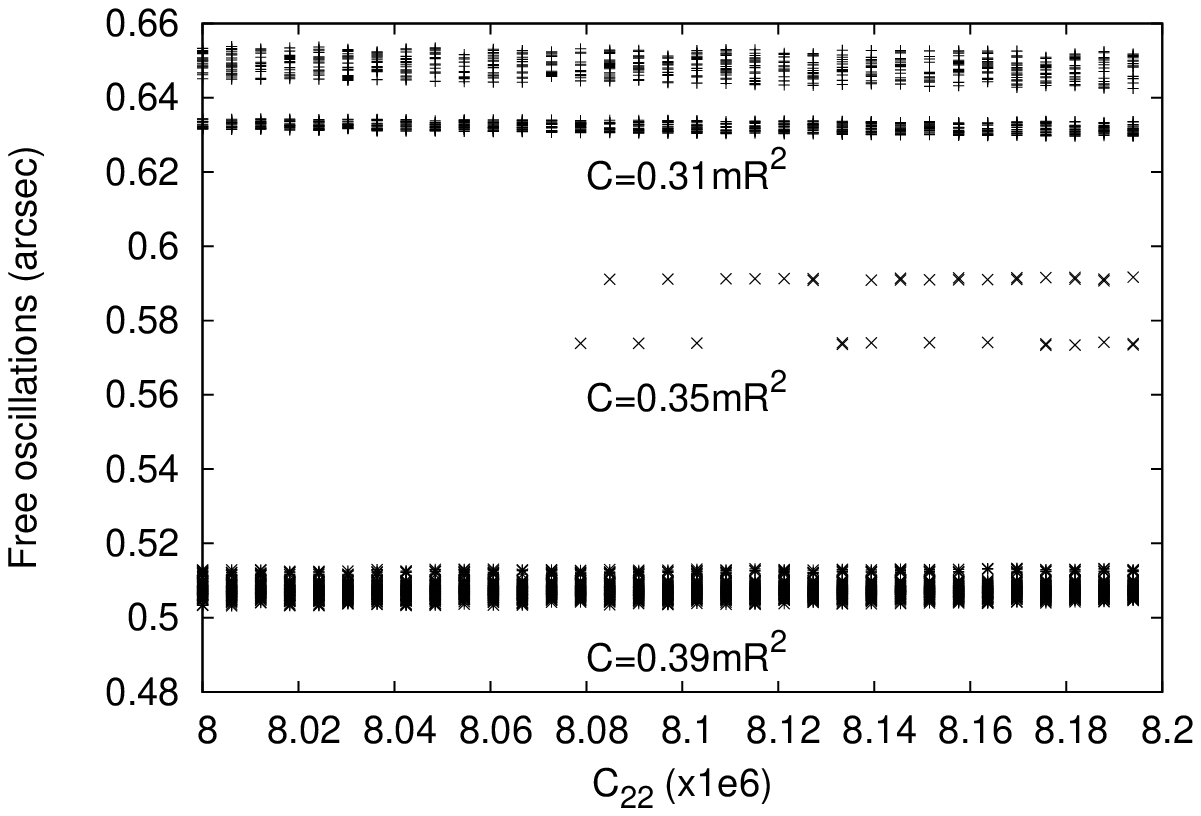} \\
with DE406 & with INPOP10a
\end{tabular}
\caption{Amplitude of the free oscillations, induced by our initial conditions. Each point is the amplitude of the free oscillations given by a frequency analysis, after a numerical integration of the Hamilton equations derived from Eq.(\ref{eq:hamil}) with our initial conditions. We considered differents sets of interior parameters ($C_{20}$, $C_{22}$, and $C$). The orbital motion of Mercury is given by DE406 (left) and INPOP10a (right).\label{fig:freeeps}}
\end{figure}

\subsection{Influence of $J_3$}

\par We failed to find an influence of $C_{30}=-J_3$ in our numerical and analytical formulae of the equilibrium obliquity. Anyway, this parameter is present in the 
full equations of the rotational dynamics of the system, and we checked its influence in plotting the average value of the obliquity $\epsilon$ with respect to $J_3$
(Fig.\ref{fig:J3}).

\begin{figure}[ht]
\centering
\includegraphics{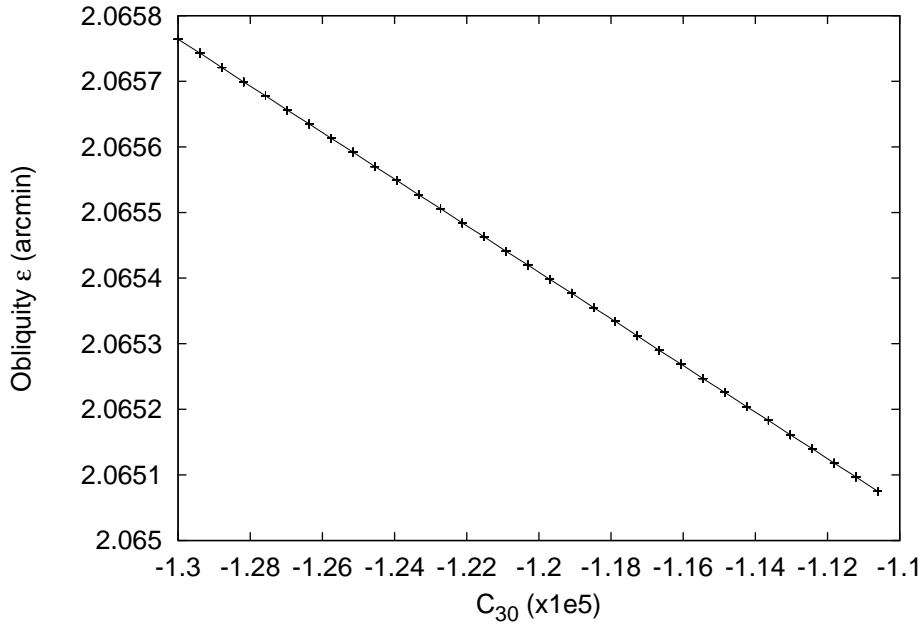}
\caption{Influence of $C_{30}$ on the mean obliquity, obtained after numerical integration of the non-averaged equations. The other interior parameters are taken in Tab.\ref{tab:smith}.\label{fig:J3}}
\end{figure}
In Fig.\ref{fig:J3} we clearly see the influence of $C_{30}$ on the obliquity $\epsilon$, which we did not see in the simple analytical nor the single 
averaged but still time dependent model. As we learned from (iii) of Sect.2.4, there is a time-dependent contribution proportional to $C_{30}$ of the
order of $eR^3/a^4$, which is affecting the action $\Sigma_3$ - thus the inertial obliquity $K$ as well as $\epsilon$ through the presence of the 
resonant argument $\sigma_3$ and the argument of the perihelion $\omega$. However, the effect is very small due to the presence of the $a^4$ in the
denominator and can safely be neglected for the same reasoning we did to explain why $J_4$ does not influence the results.

A linear fit gives 

\begin{equation}
	\label{eq:fitJ3}
	\epsilon=(-355.197C_{30}+2.06115)\quad \textrm{arcmin},
\end{equation}
so neglecting $J_3$ should induce an error of $\approx253$ mas for $J_3=1.188\times10^{-5}$. We did the same job for $J_4$ without finding any reliable influence.

\par From a theoretical point of view the difference between the averaged (but still time dependent) and original systems of equations of motion can be seen in the following way: with the average over the mean orbital longitude we neglect not only the fast periodic effects (the relevant timescale being the orbital period of Mercury), but we also set the semi-major axis of Mercury to a constant value. Thus, in the averaged system, we are unable to include all the perturbations, which affect the time evolution of the semi-major axis of the planet. In addition we needed to expand the potential into a truncated powerseries to be able to introduce the resonant argument and perform the averaging over the fast angle. Since we have limited all our expansions to the 4th order in eccentricity, an additional difference of the order $\mathcal{O}(e^5)$ is expected. Last but not least, we use the simple average rule, which is equivalent to a first order averaging in the ratio of the masses $m/M$ (the mass of Mercury over the mass of the Sun).

\subsection{The tides}

\par The shape of Mercury, if hydrostatic, is due to the influence of a centrifugal potential, due to Mercury's spin, and a tidal potential. This tidal potential can be split into a static and an oscillating potential. As a consequence the gravity field parameters $C_{20}$ and $C_{22}$ and the polar momentum of inertia $C$ should experience periodic variations alike \citep{g2004,riwlaatdr2008,vrkdr2008}:

\begin{eqnarray}
	C_{20}(t) & = & C_{20}^{static}+\frac{k_2}{2}q_te\cos l_4, \label{eq:C20rap} \\
	C_{22}(t) & = & C_{22}^{static}-\frac{k_2}{24}q_t\left(2\cos l_4-e\cos 2l_4\right), \label{eq:C22rap} \\
	C(t) & = & C^{static}-\frac{k_2}{3}q_teMR^2\cos l_4, \label{eq:Cvh}
\end{eqnarray}
where $k_2$ is the classical Love number, $e$ the eccentricity of Mercury, and $l_4=\mathcal{M}$ is the mean anomaly of Mercury. These deformations take into account the variations of the distance Sun-Mercury (proportional to the eccentricity $e$) and the 3:2 spin-orbit resonance, inducing non-synchronous rotation. This is the reason why the variation of $C_{22}$ is not proportional to the eccentricity \citep{g2004}. We also have

\begin{equation}
\label{eq:qt}
q_t=-3\frac{M}{m}\left(\frac{R}{a}\right)^3,
\end{equation}
where $M$ is the mass of the Sun, $m$ the mass of Mercury, $R$ its mean radius and $a$ its semi-major axis. With 
$e=0.2056$, $a=57,909,226.5415$ km (JPL HORIZONS), and $M/m=6.0239249\times10^6$, 
we have $q_te=-2.778369\times10^{-7}$ and $q_t=-1.351347\times10^{-6}$. The Love number $k_2$ should be between $0$ (fully inelastic rigid Mercury) and $1.5$ (fully fluid Mercury). We think, from the detection of the longitudinal librations of Mercury, that it should be closer to a rigid body than to a fluid one, so we consider $k_2=0.5$. We have in fact very few data that would help to estimate $k_2$. \citet{sswc2001} estimate it between $0.3$ and $0.45$ if Mercury has no rigid inner core, and between $0.1$ and $0.4$ if it has one. With different assumptions on the composition, \citet{rvv2009} estimate it between $0.2$ and $0.8$. The results are given in Fig.\ref{fig:tides}. We can see a peak-to-peak variation of $\approx30$ mas. We can also see the secular drift due to the variations of the orbital elements related to eccentricity and inclination, $\approx10$ mas over 20 years.

\begin{figure}[ht]
\centering
\begin{tabular}{cc}
\includegraphics[width=0.45\textwidth]{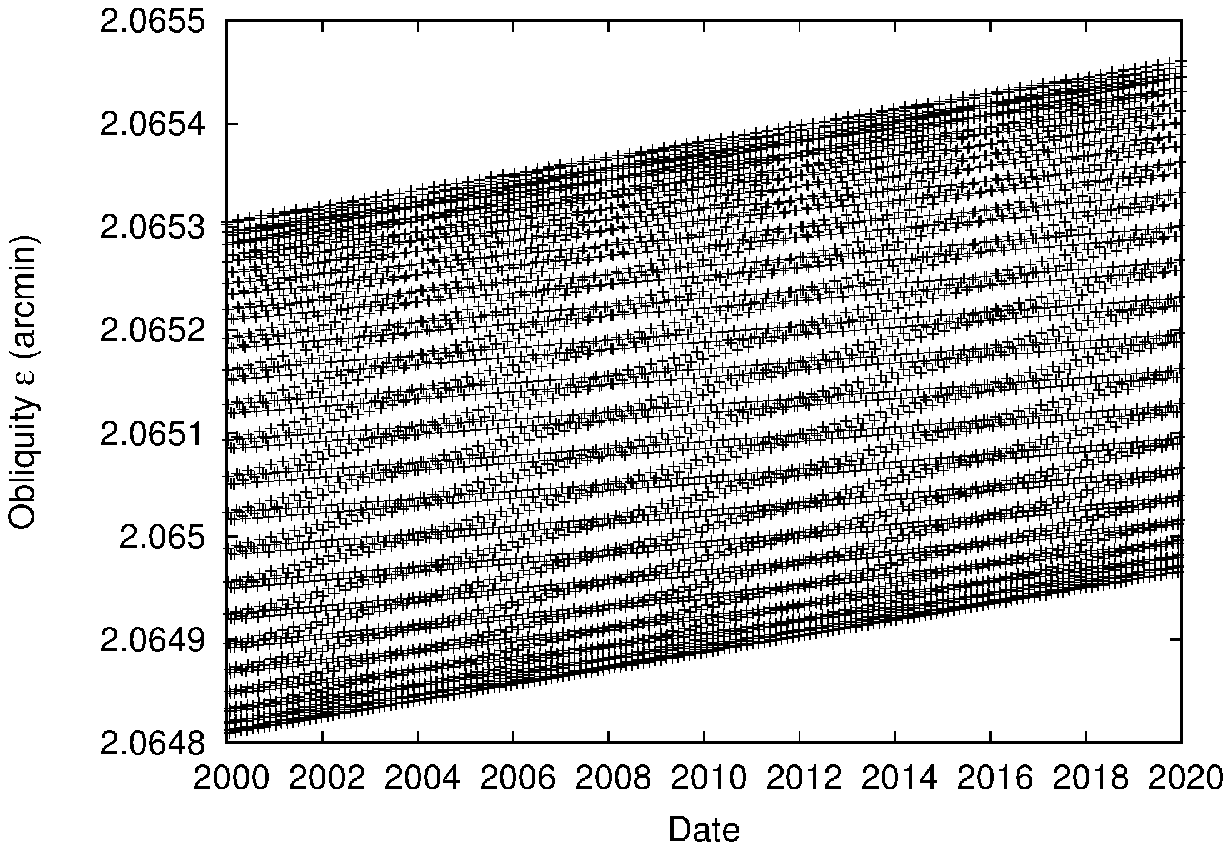} & \includegraphics[width=0.45\textwidth]{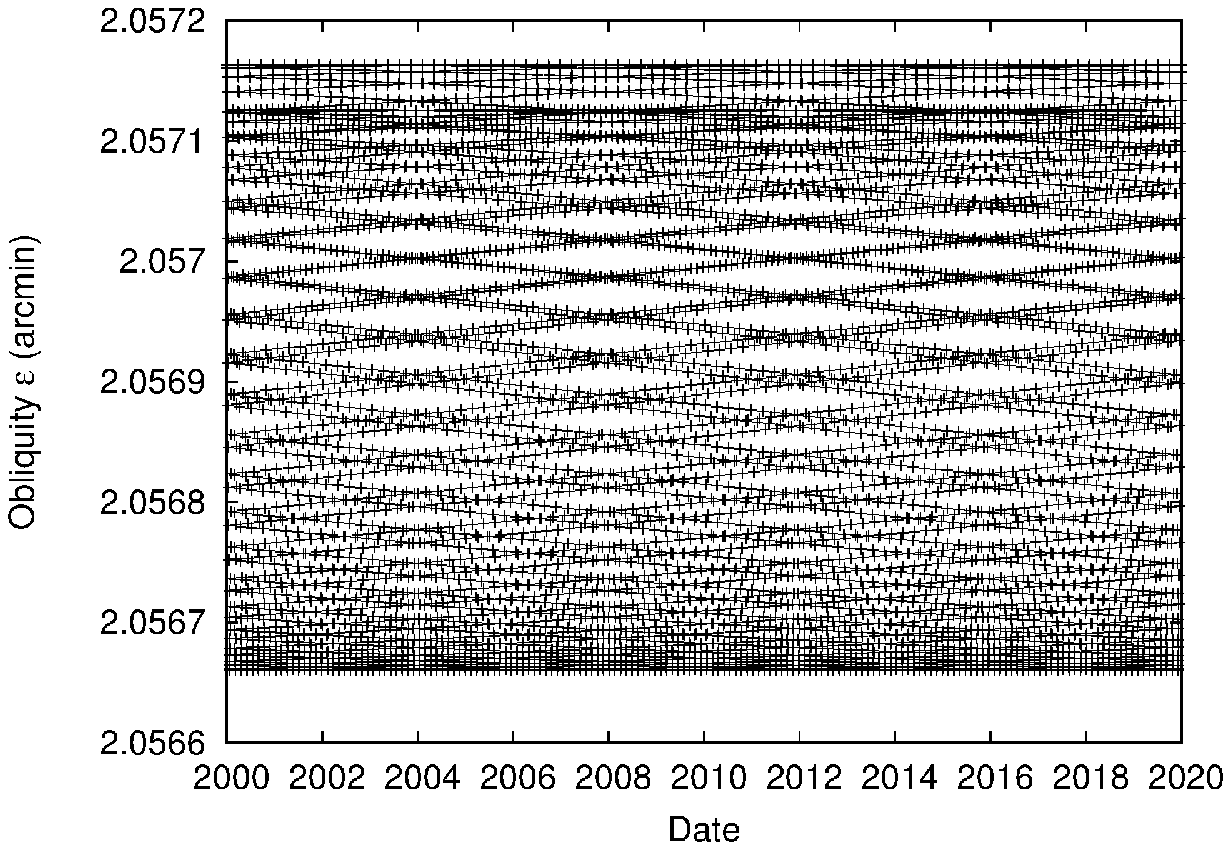}
\end{tabular}
\caption{Variations of the obliquity of Mercury due to tides, with the numerical formula (left), and the analytical one (right). The period of variation is the orbital period, i.e. 88 days.\label{fig:tides}}
\end{figure}

\subsection{Summary}

\par The influence of the different effects is gathered in the Tab.\ref{tab:effects}. The free librations that are mentioned are generated by the lack of accuracy of our initial conditions (Eq.\ref{eq:KmI} \& \ref{eq:sig3}), they do not have any physical relevance. We have made the assumptions that they have been damped to a negligible amplitude, as suggested by \citet{p2005}. The other effects are smaller than that, they all have been estimated in this study except the polar motion, coming from \citep{ndl2010}, and the amplitude of the short-period librations, from \citep{dnrl2009}. The polar motion is an oscillation of the rotation axis of Mercury about the geometrical figure axis, with a period of $175.9$ days. This motion is also plotted in (\citet{rvdb2007}, Fig.7). These numbers are very small compared to the accuracy of the determination of the obliquity, i.e. $\approx5$ arcsec.

\begin{table}[ht]
\centering
\caption{Influence on the mean obliquity of usually neglected effects. The amplitude of the free librations can be seen as an estimation of the error due to theory.\label{tab:effects}}
\begin{tabular}{l|l}
Effect & Influence on obliquity \\
\hline
Free librations & $<750$ mas \\
$C_{30}$ & $\approx250$ mas \\
Polar motion & $\approx80$ mas \citep{ndl2010} \\
Tides & $\approx30$ mas \\
Short-period librations & $<20$ mas \citep{dnrl2009} \\
Secular drift & $\approx10$ mas over 20 years \\
$C_{40}$ & negligible \\
\hline
\end{tabular}
\end{table}

\section{Inverting the obliquity of Mercury\label{sec:invert}}

\par Recently, \citet{mpshgjygpc2012} measured an obliquity of $(2.04\pm0.08)$ arcmin, and they used Peale's formula (Eq.\ref{eq:peale}) to get $C/(mR^2)=0.346\pm0.014$. In this work,
we present alternative formulae that we propose to use to invert the obliquity of \citet{mpshgjygpc2012}. We can consider that we propose 4 new formulae. From an analytical 
(Eq.\ref{EqF3}) and a numerical studies (Eq.\ref{eq:epsilon}, from Eq.\ref{eq:KmI} \& \ref{eq:sig3}) we get 2 formulae. The influence of $J_3$, that we detect only numerically with 
the full equations of the system (Fig.\ref{fig:J3} \& Eq.\ref{eq:fitJ3}) leads us to write down 2 new equations, in which the 2 obliquities given by the Eq.\ref{EqF3} \& 
\ref{eq:epsilon} are corrected by $355.197\times J_3$.

\par In considering the values of the spherical harmonics given in Tab.\ref{tab:smith}, we get the theoretical obliquity of Mercury 7 years after the date J2000.0 to be close to the mid-date of the radar observations. We considered that $C$ was the only
unknown parameter. In the analytical formula, we used the same dynamical parameters as \citet{ym2006}, i.e. $i=8.6^{\circ}$ and a regressional period of the ascending node set 
to 328 kyr. We also consider a precessional period of the pericenter $\omega$ of 128 kyr, as suggested by the JPL HORIZONS website. This number is adviced over 6,000 years. And we get

\begin{itemize}
	
\item \citet{mpshgjygpc2012}: $C/(mR^2)=0.346\pm0.014$,
	
\item Analytical formula (Eq.\ref{EqF3}): $C/(mR^2)=0.34712\pm0.01361$,
	
\item Numerical formula (Eq.\ref{eq:epsilon}): $C/(mR^2)=0.34576\pm0.01349$,
	
\item Analytical formula (Eq.\ref{EqF3}) + $J_3$: $C/(mR^2)=0.34640\pm0.01361$,
	
\item Numerical formula (Eq.\ref{eq:epsilon}) + $J_3$: $C/(mR^2)=0.34506\pm0.01348$.
	
\end{itemize}

\par All our numbers are consistent with the ones coming from Peale's formula. The polar moment of inertia of Mercury could be 
$0.345mR^2$ instead of $0.346mR^2$, this difference is very small with respect to the accuracy of the observations. Giving so many digits 
lacks of physical relevance, but is necessary to stress the tiny differences between our 4 formulae.

\section{Conclusion}

\par This study tackles the influence of usually neglected effects like the secular variations of the orbital elements, the tides and the higher order harmonics on the instantaneous obliquity of Mercury. The main goal is to invert it to get clues on the internal structure, in particular the polar inertial momentum $C$. Moreover, it gives optimized initial conditions of the orientation of the angular momentum of Mercury, at any time and for any values of the internal structure parameters, that can be directly used in numerical simulations. This is a refinement of \citep{nd2012}. These initial conditions have the advantage to be Laplace plane free, they are instead based on the ecliptic, whose definition is robust. They have been obtained thanks to averaged equations of the rotational motion, suitable for long-term studies. We hope that they will help fitting the rotation of Mercury by future experiments, like the radioscience experiment MORE in BepiColombo \citep{mrvvb2001}. A C-code implementing our formulae can be downloaded with the electronic version of this manuscript.

\par We have shown that the usually neglected effects have an influence smaller than 1 arcsec, while the observations have an accuracy of $\approx5$ arcsec. The determination of $C$ can be at the most altered from $0.346mR^2$ to $0.345mR^2$, what does not fundamentally change our understanding of the internal structure of Mercury. The size of the molten core can be obtained from the longitudinal librations of Mercury, but depends on whether we consider a rigid inner core or not \citep{vrby2012}.

\section*{Acknowledgements}

This research used resources of the Interuniversity Scientific Computing Facility located at the University of Namur, Belgium, which is supported by the F.R.S.-FNRS under convention No. 2.4617.07. It also benefited from the financial support of the contract Prodex C90253 ``ROMEO'' from BELSPO. Beno\^it Noyelles is F.R.S.-FNRS post-doctoral research fellow. The authors are indebted to the 2 reviewers, Alain Vienne and Marie Yseboodt, who pointed out some typos in formulae and whose comments significantly improved the manuscript, and to Nicolas Rambaux, who indicated them the reference to Giampieri, to properly include the tidal effects.

\section*{Appendix A}

\np We collect the various coefficients of the form $[j]_{nm}$, with
$j,n,m\in\mathbb{N}$, from Section~\ref{SPOT}. If we introduce
the notations $c_x=\cos(x)$, $s_x=\sin(x)$ they can be written in the
form:

\np At second order in $\langle V_{20}\rangle$:
\bna
[1]_{20}=3 \left(1+3 c_{2 i}\right) c_{2K} \ , \
[2]_{20}=48 c_i c_K s_i s_K \ , \
[3]_{20}=-6 c_{2 K} s_i^2 \ .
\ena

\np At second order in $\langle V_{22}\rangle$:
\bna
&&[1]_{22}=48 c_{i/2}^4 c_{K/2}^4 \ , \
[2]_{22}=12 \left(1+c_i\right) \left(1+c_K\right) s_i s_K \ , \
[3]_{22}=18s_i^2 s_K^2 \ , \\
&&[4]_{22}=48s_{i/2}^2 s_i s_{K/2}^2 s_K \ , \
[5]_{22}=48s_{i/2}^4 s_{K/2}^4 \ ,
\ena

\np in $\langle v_{22}\rangle$ and
\bna
&&[6]_{22}=-4 c_{K/2}^4 s_i^2 \ , \
[7]_{22}=4 \left(1+c_K\right) s_i c_is_K \ , \
[8]_{22}= -\left(1+3 c_{2 i}\right) s_K^2 \ , \\
&&[9]_{22}=4 c_i \left(c_K-1\right) s_i s_K \ , \
[10]_{22}=-4 s_i^2 s_{K/2}^4 \
\ena

\np in $\langle u_{22}\rangle$. At order three we have in $\langle V_{30} \rangle$:
\bna
&&[1]_{30}=-30s_{i/2}^2 s_i^2 s_K^3 \ , \
[2]_{30}=30 \left(c_i-1\right) \left(1+3 c_i\right) c_K s_i s_K^2 \ , \\
&&[3]_{30}=-\frac{3}{2}\left(13+20 c_i+15 c_{2 i}\right) 
\left(3+5 c_{2 K}\right) s_{i/2}^2 s_K \ , \\
&&[4]_{30}=-\frac{3}{4}\left(3 c_K+5 c_{3 K}\right) \left(s_i+5 s_{3 i}\right) \ , \
[5]_{30}=\frac{3}{4} c_{i/2}^2 \left(13-20 c_i+15 c_{2 i}\right) \left(s_K+5 s_{3 K}\right) \ , \\
&&[6]_{30}=30 \left(1+c_i\right) \left(3 c_i-1\right) c_K s_i s_K^2 \ , \
[7]_{30}=-15 \left(c_i-1\right) \left(1+c_i\right){}^2 s_K^3 \ .
\ena

\np The fourth order coefficients, being part of $\langle v_{40} \rangle$, turn out to be:
\bna
&&[1]_{40}=\frac{3}{64} \left(9+20 c_{2 i}+35 c_{4 i}\right) 
\left(9+20 c_{2 K}+35 c_{4 K}\right) \ , \
[2]_{40}=\frac{15}{8} \left(2 s_{2 i}+7 s_{4 i}\right) \left(2 s_{2 K}+7 s_{4 K}\right) \ , \\
&&[3]_{40}=15 \left(5+7 c_{2 i}\right) \left(5+7 c_{2 K}\right) s_i^2 s_K^2 \ , \
[4]_{40}=840 c_i c_K s_i^3 s_K^3 \ , \
[5]_{40}=105 s_i^4 s_K^4 \ .
\ena

\np The fourth order coefficients in $\langle u_{40} \rangle$ are:
\bna
&&[6]_{40}=\frac{15}{4} \left(5+7 c_{2 i}\right) \left(9+20 c_{2 K}+35 c_{4 K}\right) s_i^2 \ , \
[7]_{40}=840 s_{\frac{i}{2}}^4 s_i^2 s_K^4 \ , \\
&&[8]_{40}=6720 \left(2 c_{i/2}+c_{3 i/2}\right) c_K s_{i/2}^5 s_K^3 \ , \\
&&[9]_{40}=120 \left(9+14 c_i+7 c_{2 i}\right) \left(5+7 c_{2 K}\right) s_{i/2}^4 s_K^2 \ , \\
&&[10]_{40}=30 \left(19 c_{i/2}+7 \left(2 c_{3 i/2}+c_{5 i/2}\right)\right)
s_{i/2}^3 \left(2 s_{2 K}+7 s_{4 K}\right) \ , \\
&&[11]_{40}=-30 c_{i/2}^3 \left(19 s_{i/2}+7 \left(s_{5i/2}-2 s_{3 i/2}\right)\right)
\left(2 s_{2 K}+7 s_{4 K}\right) \ , \\
&&[12]_{40}=120 c_{i/2}^4 \left(9-14 c_i+7 c_{2 i}\right) \left(5+7 c_{2 K}\right) s_K^2 \ , \\
&&[13]_{40}=6720 c_{i/2}^5 c_K \left(s_{3 i/2}-2 s_{i/2}\right) s_K^3 \ , \
[14]_{40}=-210 \left(c_i-1\right) \left(1+c_i\right){}^3 s_K^4 \ .
\ena


\end{document}